\documentclass[subeqn,12pt,a4paper]{article}

\usepackage{amssymb,amsmath,amsfonts,amsthm, amscd, mathrsfs,helvet, mathtools}

\usepackage{bm, stmaryrd}
\usepackage{graphicx,verbatim}
\usepackage[usenames, dvipsnames]{color}
\usepackage{psfrag}
\usepackage[all]{xy}

\usepackage{cite}

\theoremstyle{plain}

\newtheorem*{theorem*}{Theorem}

\newtheorem*{proposition*}{Proposition}

\newcommand{\tensor}[1]{{\bf \underline{#1}}}

\definecolor{brightBlue}{rgb}{0,0,1}

\definecolor{Violet}{rgb}{0.47,0,1}

\DeclareMathOperator{\tr}{Tr} 
\DeclareMathOperator{\res}{res}

\def\a{\mathfrak{a}}

\def\g{\mathfrak{g}}
\def\h{\mathfrak{h}}

\def\ha{\mbox{\small $\frac{1}{2}$}}
\def\qa{\mbox{\small $\frac{1}{4}$}}

\newcommand{\bb}[1]{\llbracket #1 \rrbracket}

\def\CC{\mathbb{C}}

\def\M{\mathcal{M}}

\def\X{\mathcal{X}}
\def\Y{\mathcal{Y}}

\def\1{\tensor{1}}
\def\2{\tensor{2}}
\def\3{\tensor{3}}
\def\4{\tensor{4}}

\numberwithin{equation}{section}

\renewcommand{\L}[0]{\mathcal{L}}

\def\beq{\begin{equation}}
\def\eeq{\end{equation}}
\def\beqz{\begin{equation*}}
\def\eeqz{\end{equation*}}
\def\bea{\begin{eqnarray}}
\def\eea{\end{eqnarray}}

\def\et{\qquad\mbox{and}\qquad}
\def\fr{Faddeev-Reshetikhin }

\def\pp{\Pi}
\def\QISM{QISM} 
\def\A{A}

\def\f{\mathfrak{f}}
\def\hf{\widehat{\mathfrak{f}}}

\def\g{\mathfrak{g}}

\def\hfs{\widehat{\mathfrak{f}}^{\sigma}}
\def\hFs{\widehat{F}^{\sigma}}
\def\eqm{=}
\newcommand{\coloneqq}{=}

\begin{document}

\begin{center}
\vspace*{2em}
{\large\bf
Alleviating the non-ultralocality of coset $\sigma$-models\\
\vspace{1mm}
through a generalized Faddeev-Reshetikhin procedure}\\
\vspace{1.5em}
F. Delduc$\,{}^1$,  M. Magro$\,{}^1$, B. Vicedo$\,{}^2$

\vspace{1em}
\begingroup\itshape
{\it 1) Laboratoire de Physique, ENS Lyon
et CNRS UMR 5672, Universit\'e de Lyon,}\\
{\it 46, all\'ee d'Italie, 69364 LYON Cedex 07, France}\\
\vspace{1em}
{\it 2) Department of Mathematics, University of York,}\\
{\it Heslington, York, YO10 5DD, United Kingdom }
\par\endgroup
\vspace{1em}
\begingroup\ttfamily
Francois.Delduc@ens-lyon.fr, Marc.Magro@ens-lyon.fr, Benoit.Vicedo@gmail.com
\par\endgroup
\vspace{1.5em}
\end{center}

\paragraph{Abstract.}

The \fr procedure corresponds to a removal of the non-ultralocality of the classical $SU(2)$ principal chiral model. It is realized by defining another field theory, which has the same Lax pair and equations of motion but a different Poisson structure and Hamiltonian. Following earlier work of M. Semenov-Tian-Shansky and A. Sevostyanov, we show how it is possible to alleviate in a similar way the non-ultralocality of symmetric space $\sigma$-models. The equivalence of the equations of motion holds only at the level of the Pohlmeyer reduction of these models, which corresponds to symmetric space sine-Gordon models. This work therefore shows indirectly that symmetric space sine-Gordon models, defined by a gauged Wess-Zumino-Witten action with an integrable potential, have a mild non-ultralocality. The first step needed to construct an integrable discretization of these models is performed by determining the discrete analogue of the Poisson algebra of their Lax matrices.

\pagebreak

\section{Introduction\label{intro}}

The Quantum Inverse Scattering Method (QISM) \cite{faddtakh1979-1, Kulish:1979if, faddsklytak1980tmp1} provides a very general and successful framework for establishing and studying the quantum integrability of a broad class of $(1+1)$-dimensional quantum field theories. Yet despite its formidable success, a significant number of important quantum field theories which are believed to be quantum integrable have so far resisted its application.
Some well known examples are the principal chiral field model and the symmetric space $\sigma$-model. What distinguishes these theories is that they violate one of the key 
assumptions behind the \QISM, known as ultralocality.
In fact, the difficulty in dealing with these so called non-ultralocal 
theories is already apparent classically.
Indeed, the requirement of ultralocality classically means that 
the Poisson bracket of the Lax matrix with itself does not
depend on derivatives of the Dirac $\delta$-function. When this assumption 
fails, the computation of the Poisson bracket of the monodromy 
matrix becomes plagued with ambiguities. Attempting to 
fix these ambiguities leads to a bracket that doesn't satisfy 
the Jacobi identity \cite{Maillet:1985fn, Maillet:1985ek}. And 
although a proof of classical integrability is still possible in this 
case, the lack of a well defined Poisson bracket of monodromies 
severely hinders the introduction of an integrable lattice 
discretization for these models.

\medskip

For the $SU(2)$ principal chiral model, however, the situation is slightly better. 
Indeed, in 1986, L. Faddeev and N. Reshetikhin put forward an 
interesting proposal in \cite{Faddeev:1985qu} for circumventing 
the problem of non-ultralocality in this model. 
The first step taken in \cite{Faddeev:1985qu} was 
to replace the problematic non-ultralocal Poisson 
brackets by ultralocal ones. In doing so, 
the Hamiltonian also needs to be modified in order to reproduce the 
classical dynamics of the $SU(2)$ principal chiral model.
As a matter of fact, the new ultralocal Poisson brackets turn out to 
be degenerate, which means that they can only be used to reproduce 
a reduction of the original dynamics where the Casimirs have been 
set to constants.
Nevertheless, given this classically equivalent description of the 
model in terms of an ultralocal Poisson structure, it could be 
 quantized within the \QISM.

\medskip

It is natural to ask whether such a method can be generalized to 
other non-ultralocal models as well. Indeed, if this were possible, 
the \fr procedure may provide a consistent way of treating 
more general non-ultralocal theories and fitting them into the 
general scheme of the \QISM. The purpose of this article is to 
initiate such a program by generalizing the first steps of the 
\fr procedure to the case of symmetric space $\sigma$-models. 
Specifically, we shall propose a modification of the Poisson 
brackets and Hamiltonian of these models in the spirit of 
\cite{Faddeev:1985qu} which will lead to a well defined lattice Poisson algebra.

\medskip

The first task of determining the modified Poisson brackets 
is a kinematical one. As we shall see, unlike the case of the 
principal chiral model, it won't be possible to completely do 
away with the non-ultralocality in the Poisson brackets of 
coset $\sigma$-models. It will nevertheless be possible to 
alleviate their non-ultralocality in the following sense. It was 
shown by M. Semenov-Tian-Shansky and A. Sevostyanov in 
\cite{SemenovTianShansky:1995ha} that there exists a natural 
non-ultralocal Poisson structure on these models which, after regularization, does admit an integrable lattice discretization of the general form identified in \cite{Freidel:1991jx, Freidel:1991jv}. We shall refer to this special form 
of non-ultralocality as being mild. Our generalization of the 
\fr modification of the Poisson bracket will therefore be to 
replace the original non-ultralocal Poisson structure by this milder non-ultralocal one.

\medskip

The second task of determining the corresponding Hamiltonian is 
a dynamical one. Just as for the $SU(2)$ principal chiral model 
\cite{Faddeev:1985qu}, it turns out that the modified Poisson 
brackets are degenerate. Since the corresponding set of 
Casimirs will necessarily remain constant in time, this 
means that we can only reproduce a reduction of the 
original dynamics. Quite remarkably, the specific form 
of the Casimirs of the modified Poisson brackets leads 
naturally to performing a Pohlmeyer reduction \cite{Pohlmeyer:1975nb} of the 
symmetric space $\sigma$-model.
In other words, the equations of motion that we shall be 
able to reproduce using the modified Poisson brackets 
are precisely those of the Pohlmeyer reduction 
 of the original $\sigma$-model. 
In fact more is true.
Recall that the equations of motion obtained by this reduction 
identify with those of a symmetric space sine-Gordon theory 
\cite{Pohlmeyer:1979ch, Eichenherr:1979yw, Eichenherr:1979mx, 
Eichenherr:1979uk, D'Auria:1979tb, D'Auria:1980cx, D'Auria:1980xs, 
Zakharov:1973pp}, the Lagrangian formulation of which corresponds 
to a gauged Wess-Zumino-Witten model with an integrable potential
\cite{Bakas:1995bm} (see also \cite{Grigoriev:2007bu, Miramontes:2008wt}). 
An important result of the present work is that the canonical Poisson 
brackets associated with this latter model \cite{Bowcock:1988xr} 
precisely coincide with the alleviated non-ultralocal Poisson 
brackets of the coset $\sigma$-model. In particular, this shows 
indirectly that the non-ultralocality of gauged WZW models with 
an integrable potential is mild. 

\medskip

The plan of this article is as follows. After a short reminder of the 
problem of non-ultralocality and its formulation within the 
$R$-matrix approach \cite{SemenovTianShansky:1983ik, 
semenov_2002, Semevov2008a, Vicedo:2010qd}, in section 
\ref{sec: FR bracket} we recast the initial step of the \fr 
procedure in a general algebraic framework that enables 
a systematic and practical determination of the new Poisson 
brackets. This is achieved by first generalizing the \fr Poisson 
brackets to the principal chiral model on a generic Lie group 
 before extending these ideas to symmetric space 
$\sigma$-models. In particular, we show in both cases that 
these modified Poisson brackets are compatible with the original ones.

In section \ref{sec: Dynamics}, we discuss the dynamics of the 
coset $\sigma$-model with respect to the new Poisson brackets. 
We show that fixing the values of all the Casimirs of these  
degenerate brackets amounts to doing a Pohlmeyer reduction 
of the $\sigma$-model. As usual, the reduced 
 field equations have separate 
left and right gauge invariances which have to be partially 
fixed to the diagonal gauge invariance before the model can 
be described in the Hamiltonian framework. The resulting dynamics and Poisson 
brackets of the reduced fields are those of a gauged WZW model with 
a potential term.  Finally, we also 
write down the Lax matrix of the reduced model.

Section \ref{sec: lattice} is devoted to the first step towards
 discretization. Following \cite{SemenovTianShansky:1995ha}, 
we give an integrable lattice discretization of the Poisson brackets as in \cite{Freidel:1991jx, Freidel:1991jv}. The construction 
uses an arbitrary solution of the modified Yang Baxter equation 
on a finite dimensional Lie algebra. The Poisson bracket of the Lax matrix on the lattice are given, from which we deduce the regularized Poisson bracket of the monodromy matrix. Taking the continuum limit we then find that the former bracket correctly reproduces the Poisson bracket of the Lax matrix of the continuum theory.

Finally, some comments, a conclusion and some outlooks are 
gathered in section \ref{sec conclu}. For general notations we refer the reader to the
appendix.

\section{Generalizing the Faddeev-Reshetikhin bracket} \label{sec: FR bracket}

\subsection{Setup of the procedure}

\subsubsection{Problem with non-ultralocality} \label{sec: NUL prob}

To begin with let us briefly recall what is meant by non-ultralocality in 
classical integrable field theory and why this property leads to a severe 
obstacle 
for quantization.
Consider a classical integrable field theory on the circle whose field 
content is specified by a Lax matrix $\L(\sigma, \lambda)$. We let 
$\sigma \in S^1$ be a parameter on the circle and $\lambda \in \CC$ 
be the so called spectral parameter.
It is well known \cite{Maillet:1985ek} that a fairly general form of the 
Poisson bracket of $\L(\sigma, \lambda)$ with itself which will 
guarantee complete integrability of the theory is\footnote{For later 
convenience we have departed from the usual convention by changing 
the overall sign of $s_{\1\2}$.}
\begin{multline}
\{ \L_{\1}(\sigma, \lambda), \L_{\2}(\sigma',\mu) \} = \bigl[ r_{\1\2}(\lambda, \mu), \L_{\1}(\sigma, \lambda) + \L_{\2}(\sigma, \mu) \bigr] \delta_{\sigma \sigma'}\\
+ \bigl[ s_{\1\2}(\lambda,\mu), \L_{\1}(\sigma, \lambda) - \L_{\2}(\sigma, \mu) \bigr] \delta_{\sigma \sigma'} + 2 s_{\1\2}(\lambda,\mu) \delta'_{\sigma \sigma'}. \label{pb1}
\end{multline}
Throughout we let $\delta_{\sigma \sigma'}$ denote the Dirac $\delta$-function and set $\delta'_{\sigma \sigma'} = \partial_{\sigma} \delta_{\sigma \sigma'}$.
The theory is then said to be ultralocal if $s_{\1\2} = 0$ and non-ultralocal 
otherwise. In particular, non-ultralocal theories are characterized 
by the presence of the $\delta'_{\sigma \sigma'}$ term in \eqref{pb1}. 

A standard approach to quantizing an integrable field theory \cite{faddtakh1979-1, Kulish:1979if, faddsklytak1980tmp1} begins by introducing a lattice regularization to handle the UV divergences of the quantum theory. To discretize the classical integrable field theory, we start by breaking up the circle at a finite set of points $\sigma_n \in S^1$, $n = 1, \ldots, N$. The lattice Lax matrix $\L^n(\lambda)$ is then defined to be the parallel transporter from the site $\sigma_n$ to the next site $\sigma_{n+1}$, namely
\begin{equation*}
\L^n (\lambda) = P \overleftarrow{\exp} \int_{\sigma_n}^{\sigma_{n+1}}
\L(\sigma,\lambda) d\sigma.
\end{equation*}
By using the Leibniz rule one can reduce the computation of the Poisson bracket between $\L^n(\lambda)$ and $\L^m(\mu)$ to a double integral involving the Poisson bracket \eqref{pb1}. 
When dealing with ultralocal theories for which $s_{\1\2} = 0$, the substitution of \eqref{pb1} into this double integral is unambiguous and leads to the following ultralocal lattice algebra
\beq
\{ \L_{\1}^n(\lambda) , \L_{\2}^m(\mu)  \} = \bigl[ r_{\1\2}(\lambda, \mu) , \L_{\1}^n(\lambda) \L_{\2}^m(\mu) \bigr] \delta_{mn}, \label{lpb1}
\eeq
where $\delta_{mn}$ is the Kronecker symbol. The quantization of the lattice algebra \eqref{lpb1} constitutes the starting point for the \QISM.

However, when $s_{\1\2} \neq 0$, the presence of the $\delta'_{\sigma \sigma'}$ term in \eqref{pb1} leads to ambiguities when evaluating the double integral, the reason being that the Poisson bracket of parallel transporters is not well defined whenever any two of their end-points coincide \cite{Maillet:1985ek}. As a result, in the case of a generic non-ultralocal theory, the $\L^n(\lambda)$ do not have well defined Poisson brackets.

\subsubsection{Algebraic formulation}

In order to generalize the \fr procedure it will be essential to isolate the root of non-ultralocality. In view of this it is extremely useful to phrase the latter in a somewhat abstract setting \cite{SemenovTianShansky:1983ik}.
In this setting, the integrable field theories we shall be considering are associated with a set
\beq
(\hf, \L, R, \varphi), \label{set}
\eeq
where $\hf$ is a loop algebra, the Lax matrix $\L$ is a map from $S^1$ to $\hf$ and $R$ is an $R$-matrix, {\em i.e.} an element of $\text{End} \, \hf$ satisfying the modified classical Yang-Baxter equation (mCYBE) on $\hf$
\beq
\forall X, Y \in \hf, \qquad
[RX,RY] - R \bigl( [RX,Y] +[X,RY] \bigr) + \omega [X,Y] = 0 \label{mybm}
\eeq
with $\omega =1$. Equation \eqref{mybm} ensures (for any value
of $\omega$) that the so called $R$-bracket, defined by
\beq
[X,Y]_R \eqm [RX,Y]+[X,RY] \label{rbra}
\eeq
is a Lie bracket on $\hf$.
The last input in \eqref{set} is a formal Laurent series $\varphi(\lambda) \in \CC(\!( \lambda )\!)$, called the twist, specifying an inner product on $\hf$. Fixing a non singular invariant inner product $\langle \cdot, \cdot \rangle$ on $\f$, the latter is defined in terms of $\varphi$ by taking the following residue
\begin{equation} \label{twisted ip}
(X, Y)_{\varphi} \coloneqq \text{res}_{\lambda = 0} \, d\lambda \, \varphi(\lambda) \langle X(\lambda), Y(\lambda) \rangle,
\end{equation}
for any $X, Y \in \hf$. This is sometimes referred to as the twisted inner product on $\hf$.

The Poisson brackets \eqref{pb1} of the corresponding integrable field theory can now be expressed in terms of the data \eqref{set} as follows.
We equip the space $C^{\infty}(S^1, \hf)$, to which $\L$ belongs, with the following inner product and cocycle,
\begin{equation} \label{ip and cocycle}
(\!( \X, \Y )\!)_{\varphi} \coloneqq \int_{S^1} d\sigma (\X(\sigma),
\Y(\sigma))_{\varphi}, \qquad \omega_{\varphi}( \X, \Y ) \coloneqq
\int_{S^1} d\sigma (\X(\sigma), \partial_{\sigma} \Y(\sigma))_{\varphi}.
\end{equation}
The Poisson bracket between any two functions $f, g$ of the Lax matrix can then be written as
\begin{equation} \label{lax PB}
\{ f, g \} (\L) = \bigl(\!\bigl( \L, [R\, d_{\varphi} f, d_{\varphi} g] +
[d_{\varphi} f, R\, d_{\varphi} g] \bigr)\!\bigr)_{\varphi} + \bigl(\omega_{\varphi}
(R\, d_{\varphi} f, d_{\varphi} g) + \omega_{\varphi}(d_{\varphi} f, R\, d_{\varphi} g)
\bigr).
\end{equation}
The subscript $\varphi$ on the differential is used to indicate that $d_{\varphi} f(\L)$ is defined relative to the inner product \eqref{ip and cocycle}, in other words
\begin{equation*} \label{der f}
\big(\!\big( \X, d_{\varphi} f(\L) \big)\!\big)_{\varphi} \coloneqq \left. \frac{d}{dt} \right|_{t = 0} f(\L + t \X).
\end{equation*}
This Poisson bracket is merely the Kostant-Kirillov bracket on the central extension of $C^{\infty}(S^1, \hf)$, defined by the cocycle $\omega_{\varphi}$, associated with the $R$-bracket \eqref{rbra}. To bring it to a more recognizable form comparable with \eqref{pb1} we restrict attention to linear functions $f, g$ of $\L \in C^{\infty}(S^1, \hf)$ and use tensor notation. Letting $R^{\ast}$ denote the adjoint of $R$ with respect to \eqref{twisted ip} one finds
\begin{equation} \label{lax PB tensor}
\{ \L_{\1}(\sigma), \L_{\2}(\sigma') \} = [R_{\1\2}, \L_{\1}(\sigma)] \delta_{\sigma \sigma'} - [R^{\ast}_{\1\2}, \L_{\2}(\sigma)] \delta_{\sigma \sigma'} + (R_{\1\2} + R^{\ast}_{\1\2}) \delta'_{\sigma \sigma'}.
\end{equation}
We refer the reader for instance to \cite{Vicedo:2010qd} for details but simply note here that the term in $\delta'_{\sigma \sigma'}$ comes precisely from the cocycle $\omega_{\varphi}$. The Poisson bracket \eqref{lax PB tensor} is then identified with \eqref{pb1} if we define the matrices $r_{\1\2}$ and $s_{\1\2}$ to be the kernels of the skew-symmetric and symmetric parts of $R$ respectively,
\beq
r = \frac{1}{2}(R - R^\ast) \et s = \frac{1}{2} (R + R^\ast). \label{rsR}
\eeq
It is apparent from \eqref{rsR} that ultralocal theories correspond to the situation where the $R$-matrix is skew-symmetric with respect to the inner product, \emph{i.e.} $R^\ast =-R$.

As we will see below, on an abstract level the procedure of \cite{Faddeev:1985qu} consists in keeping the same loop algebra, Lax matrix and $R$-matrix but changing the inner product in such a way that $R^\ast =-R$ with respect to the new inner product. 
This has the desired effect of replacing the problematic non-ultralocal Poisson bracket of the Lax matrix of the $SU(2)$ principal chiral model by an ultralocal one. When expressed in terms of the dynamical fields, this latter Poisson bracket corresponds precisely to the modified bracket of \cite{Faddeev:1985qu}. In fact, we will show this more generally by working with the principal chiral model on a generic Lie group. We then generalize these ideas to symmetric space $\sigma$-models.

\subsection{Principal chiral model}

\subsubsection{Original bracket} \label{sec: PCM orig}

We start by identifying the set $(\hf, \L, R, \varphi)$ in the case of the principal chiral model on a Lie group $F$. The first element is simply the loop algebra $\hf \coloneqq \f \otimes \mathbb{C}(\!( \lambda )\!)$ of formal Laurent series with coefficients in $\f = \text{Lie}(F)$. In terms of the usual components $(j_0, j_1)$ of the current $j$ taking values in $\f$, the Lax matrix of the principal chiral model is given by
\begin{equation} \label{PCM lax}
\L \coloneqq \frac{1}{1 - \lambda^2} (j_1 + \lambda \, j_0).
\end{equation}
Next, the $R$-matrix is defined by choosing a pair of complementary subalgebras of $\hf$. In the obvious notation we let
\begin{equation*}
\hf_{\geq 0} = \f \otimes \CC\bb{\lambda}, \qquad \hf_{< 0} = \f \otimes \lambda^{-1} \CC[\lambda^{-1}],
\end{equation*}
and similarly for $\hf_{> 0}$ and $\hf_{\leq 0}$. Let $\pi_{\geq 0}$, $\pi_{< 0}$, $\pi_{> 0}$ and $\pi_{\leq 0}$ be the projections of $\hf$ onto these respective subalgebras. For later purposes let us also introduce the projection $\pi_0$ onto the constant subalgebra $\f \subset \hf$. The standard $R$-matrix can now be defined as
\begin{equation} \label{PCM R-matrix}
R = \pi_{\geq 0} - \pi_{< 0}.
\end{equation}
Finally, the inner product \eqref{twisted ip} on $\hf$ is given by the following choice of twist
\begin{equation} \label{PCM twist}
\varphi(\lambda) \coloneqq 2 \left(1 - \frac{1}{\lambda^2} \right).
\end{equation}
As a result of this twist, the $R$-matrix \eqref{PCM R-matrix} is not skew-symmetric. Indeed, the adjoint $R^{\ast}$ of $R$ can be computed explicitly as
\begin{align*}
 \text{res}_{\lambda = 0} \, d\lambda \, \varphi(\lambda) \bigl\langle R \bigl( X(\lambda) \bigr), Y(\lambda) \bigr\rangle
&= \text{res}_{\lambda = 0} \, d\lambda \, \bigl\langle \pi_{\geq 0}\bigl(X(\lambda)\bigr)  - \pi_{< 0} \bigl(X(\lambda)\bigr), \varphi(\lambda) Y(\lambda) \bigr\rangle, \\
&= \text{res}_{\lambda = 0} \, d\lambda \, \bigl\langle X(\lambda), \pi_{< 0} \bigl( \varphi(\lambda) Y(\lambda)\bigr) - \pi_{\geq 0}\bigl( \varphi(\lambda) Y(\lambda)\bigr) \bigr\rangle, \\
&= - \text{res}_{\lambda = 0} \, d\lambda \, \varphi(\lambda) \bigl\langle X(\lambda), \varphi(\lambda)^{-1} R\bigl(\varphi(\lambda) Y(\lambda)\bigr) \bigr\rangle,
\end{align*}
from which we deduce that
\beq
R^{\ast} = - \tilde{\varphi}^{-1}  \circ R  \circ \tilde{\varphi}, \label{genrad}
\eeq
where $\tilde{\varphi}$ denotes the multiplication by $\varphi(\lambda)$.

\medskip

The expression \eqref{lax PB} is very useful if one wants to derive the Poisson brackets of the currents from those of the Lax matrix. Indeed, consider $x \in \f$. Then $\lambda x$ is in $\hf$ and we have
\beqz
(\!( \L, - \ha \lambda x \cdot \delta_{\sigma} )\!)_{\varphi} =
 - \int_{S^1} d\sigma'  \, \delta_{\sigma \sigma'} \, \text{res}_{\lambda = 0} \, d\lambda \,\frac{\varphi(\lambda)}{1-\lambda^2}  \bigl\langle j_1(\sigma') + \lambda j_0(\sigma'), \ha \lambda x \bigr\rangle = \langle j_1(\sigma), x \rangle.
\eeqz
We define then two linear functions for any $x \in \f$ as
\begin{equation} \label{linear functions}
j^0_{\sigma, x} : \L \mapsto (\!( \L, - \ha x \cdot
\delta_{\sigma} )\!)_{\varphi} = \langle j_0(\sigma), x \rangle, \qquad
j^1_{\sigma, x} : \L \mapsto (\!( \L, - \ha \lambda x \cdot
\delta_{\sigma} )\!)_{\varphi} = \langle j_1(\sigma), x \rangle.
\end{equation}
In particular, since these functions are linear we have
\beqz
d_{\varphi} j^0_{\sigma, x} = - \ha x \cdot \delta_{\sigma} \et
d_{\varphi} j^1_{\sigma, x} = - \ha \lambda x \cdot \delta_{\sigma}.
\eeqz
We can now extract the Poisson brackets of the fields 
$j_0$, $j_1$ from \eqref{lax PB}. For instance,
\begin{align*}
\langle \{ j_{0\1}(\sigma), j_{1\2}(\sigma') \}, x_{\1} y_{\2}
\rangle_{\1\2}  &=   \{ j^0_{\sigma, x}, j^1_{\sigma', y} \} (\L),\\
&= \qa \bigl( \L, [R(x) , \lambda  y] + [ x, R (\lambda  y)]
\bigr)_{\varphi} \, \delta_{\sigma \sigma'}\\
&\qquad\qquad\qquad + \qa \Bigl(\bigl(R( x) , \lambda y\bigr)_{\varphi}
+ \bigl(x, R (\lambda  y)\bigr)_{\varphi} \Bigr) \, \delta'_{\sigma \sigma'}\\
&= \ha \bigl( \L, \lambda  [x , y] \bigr)_{\varphi} \, \delta_{\sigma \sigma'}
+ \ha \bigl( x , \lambda  y\bigr)_{\varphi} \, \delta'_{\sigma \sigma'}\\
&= - \langle j_1(\sigma), [x , y] \rangle \, \delta_{\sigma \sigma'}
- \langle x , y \rangle \, \delta'_{\sigma \sigma'}.
\end{align*}
The bracket $\{ j_0, j_1 \}$ follows from this computation since
 $x, y \in \f$ are arbitrary. The remaining brackets
$\{ j_0, j_0 \}$ and $\{ j_1, j_1 \}$ are obtained in a similar way
and altogether we recover the Poisson brackets of the
principal chiral model
\begin{subequations} \label{standard PCM brackets}
\begin{align}
\{ j_{0\1}(\sigma), j_{0\2}(\sigma') \} &= [C_{\1\2}, j_{0\2}(\sigma)]
 \delta_{\sigma \sigma'},\\
\{ j_{0\1}(\sigma), j_{1\2}(\sigma') \} &= [C_{\1\2}, j_{1\2}(\sigma)]
\delta_{\sigma \sigma'} - C_{\1\2} \delta'_{\sigma \sigma'},\\
\{ j_{1\1}(\sigma), j_{1\2}(\sigma') \} &= 0.
\end{align}
\end{subequations}

\subsubsection{Ultralocal bracket}
\label{se122}
The non-ultralocality of the model is a consequence of the fact that its $R$-matrix \eqref{PCM R-matrix} is not skew with respect to the inner product \eqref{twisted ip} with the twist \eqref{PCM twist}. Indeed, $R$ does not commute with $\tilde{\varphi}$ and therefore
\beqz
R^{\ast} = - \tilde{\varphi}^{-1}  \circ R  \circ \tilde{\varphi} \neq -R.
\eeqz
However, it is also clear that $R$ would be skew if we had used the twist function $\varphi' = 1$ instead of \eqref{PCM twist}. This corresponds to choosing the rational inner product on $\hf$
\begin{equation} \label{rational ip}
(X, Y)_{\rm rat} \coloneqq \res_{\lambda = 0} d\lambda \, \langle X(\lambda), Y(\lambda) \rangle.
\end{equation}
Therefore, a natural prescription for obtaining an ultralocal model
is simply to replace the twisted inner product \eqref{twisted ip} by the rational inner
product \eqref{rational ip} while keeping everything else identical.
In particular we don't modify the underlying loop algebra $\hf$, we
keep the same Lax matrix $\L$ and we don't even change the
$R$-matrix!

Since the $R$-matrix \eqref{PCM R-matrix} is skew-symmetric with respect to the inner product \eqref{rational ip}, the last term in \eqref{lax PB} vanishes, leaving
\begin{equation} \label{lax PB FR}
\{ f, g \}' (\L) = (\!( \L, [R\, d_1 f, d_1 g] + [d_1 f, R\, d_1 g] )\!)_{\text{rat}}.
\end{equation}
To find the resulting Poisson bracket expressed in terms of the fields $j_0$ and $j_1$ themselves we must first find how to extract these from the Lax connection. The analogues of the linear functions \eqref{linear functions} in the present case read
\begin{equation*}
j'^0_{\sigma, x} : \L \mapsto (\!( \L, \lambda^{-2} x \cdot \delta_{\sigma} )\!)_{\text{rat}} = \langle j_0, x \rangle, \qquad
j'^1_{\sigma, x} : \L \mapsto (\!( \L, \lambda^{-1} x \cdot \delta_{\sigma} )\!)_{\text{rat}} = \langle j_1, x \rangle.
\end{equation*}
It is now straightforward to compute for example
\begin{equation*}
\langle \{ j_{0 \1}(\sigma), j_{1 \2}(\sigma') \}' , x_{\1} y_{\2} \rangle_{\1\2}
=  \{ j'^0_{\sigma, x}, j'^1_{\sigma', y} \}' (\L)
= - 2 \langle j_1(\sigma), [x, y] \rangle \, \delta_{\sigma \sigma'}.
\end{equation*}
The other brackets between the fields can be computed similarly and the result reads
\begin{subequations} \label{FR PCM brackets}
\begin{align}
\{ j_{0 \1}(\sigma), j_{0 \2}(\sigma') \}' &= 2 [C_{\1\2}, j_{0\2}(\sigma)]
\, \delta_{\sigma \sigma'},\\
\{ j_{0 \1}(\sigma), j_{1 \2}(\sigma') \}' &= 2 [C_{\1\2}, j_{1\2}(\sigma)]
\, \delta_{\sigma \sigma'},\\
\{ j_{1 \1}(\sigma), j_{1 \2}(\sigma') \}' &= 2 [C_{\1\2}, j_{0\2}(\sigma)]
\, \delta_{\sigma \sigma'}.
\end{align}
\end{subequations}
Up to an irrelevant overall factor of $2$, this is exactly the modified Poisson structure of Faddeev-Reshetikhin introduced in \cite{Faddeev:1985qu} in the context of the $SU(2)$ principal chiral model. Here we have rederived the same brackets for an arbitrary Lie group $F$ by following the simple prescription
\begin{equation} \label{algebraic FR}
(\hf, \L, R, \varphi)  \quad \longrightarrow \quad (\hf, \L, R, 1).
\end{equation}

\subsection{Symmetric space $\sigma$-model}

\subsubsection{Original bracket}

The phase-space of a $\mathbb{Z}_2$-graded coset
$\sigma$-model is parametrized by the two gradings of
the field $A = A^{(0)} + A^{(1)}$ and its canonically
conjugate momentum $P = P^{(0)} + P^{(1)}$, where
the gradings of the Lie algebra $\text{Lie}(F) = \f = \f^{(0)}
\oplus \f^{(1)}$ are defined as the eigenspaces of an involution
$\sigma : \f \to \f$ with $\sigma^2 = \text{id}$. The Lax matrix
of the model reads \cite{Vicedo:2009sn}
\begin{equation}
\L \coloneqq A^{(0)} + \ha (\lambda^{-1} + \lambda) A^{(1)} +
\ha (1 - \lambda^2) \pp^{(0)} + \ha (\lambda^{-1} - \lambda) \pp^{(1)}, \label{laxfg}
\end{equation}
where $\pp \eqm  \partial_\sigma P - [A, P]$. To describe the algebraic
structure of the model we introduce the
twisted loop algebra $\hfs$. This is the subalgebra of the
loop algebra $\hf$ consisting of elements $X(\lambda)$ which
are invariant under the automorphism $\hat{\sigma}(X)(\lambda)
\coloneqq \sigma[X(-\lambda)]$. Concretely, we have\footnote{The
number of terms with $n$ negative is arbitrary but finite. Also, in this formula and in the rest of this article, an integer between parenthesis is only considered modulo 2. Thus, $(n)$ is $(0)$ or $(1)$, depending on the parity of $n$.}
\begin{equation*}
\hfs = \bigoplus_{n} \f^{({n})} \cdot \lambda^n.
\end{equation*}
We denote by $\hfs_{< 0} = \hf_{< 0} \cap \hfs$ the subalgebra of $\hfs$ for which the direct sum is restricted to $n < 0$, and similarly for $\hfs_{\leq 0}$, $\hfs_{> 0}$, $\hfs_{\geq 0}$. By abuse of notation we shall denote the restriction of the respective projections $\pi_{< 0}$, $\pi_{\leq 0}$, $\pi_{> 0}$ and $\pi_{\geq 0}$ to $\hfs$ by the same symbol. The twisted loop algebra inherits the decomposition $\hfs = \hfs_{< 0} \dotplus \hfs_{\geq 0}$ from $\hf$ so that we may use the $R$-matrix as in \eqref{PCM R-matrix}. We also endow $\hfs$ with a twisted inner product \cite{Sevostyanov:1995hd}
\begin{equation*}
(X, Y)_{\varphi} \coloneqq \text{res}_{\lambda = 0} \frac{d\lambda}{\lambda}
\, \phi(\lambda) \, \langle X(\lambda), Y(\lambda) \rangle, \qquad
 \phi(\lambda) \coloneqq \frac{4 \lambda^2}{(1 - \lambda^2)^2}.
\end{equation*}
Notice that we have extracted an explicit factor of $\lambda^{-1}$
from the twist function $\varphi(\lambda) = \lambda^{-1} \phi(\lambda)$ so that the remaining twist $\phi(\lambda)$ is invariant under $\lambda \mapsto - \lambda$.

To extract the Poisson brackets of the fundamental fields
$A^{(i)}$ and $\pp^{(i)}$ from \eqref{lax PB} in this case we consider the following linear functionals
\begin{align*}
a^{(1)}_{\sigma, x} : \L \mapsto (\!( \L, \qa (\lambda^{-3} - \lambda^{-1}) x^{(1)} \cdot
\delta_{\sigma} )\!)_{\varphi} &= \langle A^{(1)}(\sigma), x^{(1)} \rangle,\\
\pi^{(1)}_{\sigma, x} : \L \mapsto (\!( \L, \qa (3 \lambda^{-1} - \lambda^{-3}) x^{(1)} \cdot
\delta_{\sigma} )\!)_{\varphi} &= \langle \pp^{(1)}(\sigma), x^{(1)} \rangle,\\
a^{(0)}_{\sigma, x} : \L \mapsto (\!( \L, \qa (\lambda^{-4} - \lambda^{-2}) x^{(0)} \cdot
\delta_{\sigma} )\!)_{\varphi} &= \langle A^{(0)}(\sigma), x^{(0)} \rangle,\\
\pi^{(0)}_{\sigma, x} : \L \mapsto (\!( \L, \ha (2 \lambda^{-2} - \lambda^{-4}) x^{(0)} \cdot
\delta_{\sigma} )\!)_{\varphi} &= \langle \pp^{(0)}(\sigma), x^{(0)} \rangle.
\end{align*}
In terms of these we can compute for instance,
\begin{align*}
\langle \{ A^{(0)}_{\1}(\sigma), \pp^{(0)}_{\2}(\sigma') \},
x^{(0)}_{\1} y^{(0)}_{\2} \rangle_{\1\2} &= \{ a^{(0)}_{\sigma, x},
\pi^{(0)}_{\sigma', y} \} (\L)\\
&= - \langle A^{(0)}(\sigma), [x^{(0)}, y^{(0)}] \rangle \,
\delta_{\sigma \sigma'} - \langle x^{(0)}, y^{(0)} \rangle \,
\delta'_{\sigma \sigma'}.
\end{align*}
Performing similar calculations, altogether we find exactly the Poisson
brackets of the symmetric space $\sigma$-model, namely
\begin{subequations} \label{standard Z2 brackets}
\begin{align}
\{ A^{(i)}_{ \1}(\sigma), A^{(j)}_{ \2}(\sigma') \} &= 0,\\
\{ A^{(i)}_{ \1}(\sigma), \pp^{(j)}_{\2}(\sigma') \} &=
[C^{(ii)}_{\1\2}, A^{(i + j)}_{\2}(\sigma)] \, \delta_{\sigma \sigma'} -
\delta_{ij} \, C^{(ii)}_{\1\2} \delta'_{\sigma \sigma'},\\
\{ \pp^{(i)}_{\1}(\sigma), \pp^{(j)}_{\2}(\sigma') \}
&= [C^{(ii)}_{\1\2}, \pp^{(i + j)}_{\2}(\sigma)] \, \delta_{\sigma
\sigma'}.
\end{align}
\end{subequations}

\subsubsection{Mildly non-ultralocal bracket}

Guided by our algebraic reformulation in \eqref{algebraic FR} of
the Faddeev-Reshetikhin modification of the Poisson bracket,
we would like to obtain a similar prescription in the case at
hand for symmetric space $\sigma$-model.

Since the $R$-matrix of the model is of the same form as in the principal chiral model case, naively
 one might try to replace the twisted
inner product by the rational inner product also in the present case.
However, the latter vanishes identically on
$\hfs$. Indeed, the quantity $\langle X(\lambda) Y(\lambda)\rangle$
in (\ref{rational ip}) is a function of $\lambda^2$ and therefore
has vanishing residue.
We are thus forced to use the simplest non-degenerate inner product
on $\hfs$, which is the trigonometric one,
\begin{equation} \label{trig ip}
(X, Y)_{\rm trig} \coloneqq \res_{\lambda = 0} d\lambda \, \lambda^{-1}
\langle X(\lambda), Y(\lambda) \rangle.
\end{equation}
This corresponds to the choice of twist function $\varphi'(\lambda) = \lambda^{-1}$. So our prescription for modifying the Poisson bracket is simply
\begin{equation*}
(\hfs, \L, R, \lambda^{-1} \phi)  \quad \longrightarrow \quad (\hfs, \L, R, \lambda^{-1}).
\end{equation*}
In particular, we keep the same $R$-matrix which isn't skew-symmetric with respect to the trigonometric inner product. Therefore the new Poisson bracket is still non-ultralocal! This is in sharp contrast with the Faddeev-Reshetikhin prescription for principal chiral models. However, as we will explain in section \ref{sec: lattice}, it turns out that although the new bracket is still non-ultralocal, it is not plagued with the same problems as the original non-ultralocal bracket. In particular, this better behaved non-ultralocal Poisson structure admits a lattice discretization. We refer to this special type of non-ultralocality as being mild, to be defined more precisely in section \ref{sec: lattice}. This observation leads to a natural generalization of the Faddeev-Reshetikhin procedure.

To extract the Poisson brackets of the fields $A^{(i)}$ and $\pp^{(i)}$ we proceed as usual and
consider the linear functionals
\begin{align*}
a'^{(1)}_{\sigma, x} : \L \mapsto (\!( \L, (\lambda + \lambda^{-1}) x^{(1)} \cdot \delta_{\sigma} )\!)_{\rm trig} &= \langle A^{(1)}(\sigma), x^{(1)} \rangle,\\
\pi'^{(1)}_{\sigma, x} : \L \mapsto (\!( \L, (\lambda - \lambda^{-1}) x^{(1)} \cdot \delta_{\sigma} )\!)_{\rm trig} &= \langle \pp^{(1)}(\sigma), x^{(1)} \rangle,\\
a'^{(0)}_{\sigma, x} : \L \mapsto (\!( \L, (1 + \lambda^{-2}) x^{(0)} \cdot \delta_{\sigma} )\!)_{\rm trig} &= \langle A^{(0)}(\sigma), x^{(0)} \rangle,\\
\pi'^{(0)}_{\sigma, x} : \L \mapsto (\!( \L, - 2 \lambda^{-2} x^{(0)} \cdot \delta_{\sigma} )\!)_{\rm trig} &= \langle \pp^{(0)}(\sigma), x^{(0)} \rangle.
\end{align*}
Using these expressions we can explicitly compute the
Poisson brackets between the various fields $A^{(i)}$ and
$\pp^{(i)}$. For instance,
\begin{align*}
\langle \{ A^{(0)}_{ \1}(\sigma), A^{(0)}_{ \2}(\sigma') \}', x^{(0)}_{\1} y^{(0)}_{\2} \rangle_{\1\2} &= \{ a'^{(0)}_{\sigma, x}, a'^{(0)}_{\sigma', y} \}' (\L)\\
&= \langle 2 A^{(0)}(\sigma) + \pp^{(0)}(\sigma), [x^{(0)}, y^{(0)}] \rangle \, \delta_{\sigma \sigma'} + 2 \langle x^{(0)}, y^{(0)} \rangle \, \delta'_{\sigma \sigma'}.
\end{align*}
Similarly all the other Poisson brackets can be computed. The final result for all the brackets reads
\begin{subequations} \label{FR Z2}
\begin{align}
\label{FR Z2 a}
\{ A^{(0)}_{ \1}(\sigma), A^{(0)}_{ \2}(\sigma') \}' &=
- [C^{(00)}_{\1\2}, 2 A^{(0)}_{\2}(\sigma) +
\pp^{(0)}_{\2}(\sigma)] \delta_{\sigma \sigma'} +
2 C^{(00)}_{\1\2} \delta'_{\sigma \sigma'},\\
\label{FR Z2 b}
\{ A^{(0)}_{ \1}(\sigma), A^{(1)}_{ \2}(\sigma') \}' &=
- [C^{(00)}_{\1\2}, A^{(1)}_{\2}(\sigma) + \pp^{(1)}_{\2}(\sigma)]
\delta_{\sigma \sigma'},\\
\label{FR Z2 d}
\{ A^{(0)}_{ \1}(\sigma), \pp^{(0)}_{\2}(\sigma') \}' &= 0,\\
\label{FR Z2 e}
\{ A^{(0)}_{ \1}(\sigma), \pp^{(1)}_{\2}(\sigma') \}' &=
- [C^{(00)}_{\1\2}, A^{(1)}_{\2}(\sigma) + \pp^{(1)}_{\2}(\sigma)]
\delta_{\sigma \sigma'},\\
\label{FR Z2 c}
\{ A^{(1)}_{ \1}(\sigma), A^{(1)}_{ \2}(\sigma') \}' &=
- [C^{(11)}_{\1\2}, \pp^{(0)}_{\2}(\sigma)] \delta_{\sigma \sigma'},\\
\label{FR Z2 f}
\{ A^{(1)}_{ \1}(\sigma), \pp^{(0)}_{\2}(\sigma') \}' &= 0,\\
\label{FR Z2 g}
\{ A^{(1)}_{ \1}(\sigma), \pp^{(1)}_{\2}(\sigma') \}' &=
[C^{(11)}_{\1\2}, \pp^{(0)}_{\2}(\sigma)] \delta_{\sigma \sigma'},\\
\label{FR Z2 h}
\{ \pp^{(0)}_{\1}(\sigma), \pp^{(0)}_{\2}(\sigma') \}'
 &= 0,\\
\label{FR Z2 i}
\{ \pp^{(0)}_{\1}(\sigma), \pp^{(1)}_{\2}(\sigma') \}'
&= 0,\\
\label{FR Z2 j}
\{ \pp^{(1)}_{\1}(\sigma), \pp^{(1)}_{\2}(\sigma') \}'
 &= - [C^{(11)}_{\1\2}, \pp^{(0)}_{\2}(\sigma)] \,
\delta_{\sigma \sigma'}.
\end{align}
\end{subequations}
Note that only the Poisson bracket of the field $A^{(0)}$ with itself is non-ultralocal.

\subsection{Compatibility of the Poisson brackets}

It turns out that for both the principal chiral model and symmetric space $\sigma$-model considered above, the new Poisson bracket $\{ \cdot, \cdot \}'$ is compatible with the original one $\{ \cdot, \cdot \}$, in the sense that any linear combination $u \, \{ \cdot, \cdot \}' + v \, \{ \cdot, \cdot \}$ is also a Poisson bracket.

To explain the origin of this property, we restrict ourselves to the case of the principal chiral model. We have shown that the original Poisson bracket is associated with the data $(\hf, \L, R, \varphi)$, with $\varphi$ given by \eqref{PCM twist}, through the formula \eqref{lax PB}.
However, an equivalent set of data producing the same bracket through \eqref{lax PB} is $(\hf, \L, R \circ \tilde{\varphi}^{-1}, 1)$.
Indeed,
the Poisson bracket \eqref{lax PB} can be rewritten purely in terms of the rational inner product and corresponding cocycle provided we use the twisted $R$-matrix $R_{\varphi^{-1}} = R \circ \tilde{\varphi}^{-1}$ instead of $R$. Note that $R_{\varphi^{-1}}$ also satisfies the mCYBE \eqref{mybm} but with $\omega = \varphi^{-2}$. Explicitly we have
\begin{equation*}
\{ f, g \} (\L) = (\!( \L, [ R_{\varphi^{-1}} d_1 f, d_1 g] + [d_1 f, R_{\varphi^{-1}} d_1 g] )\!)_{\rm rat} + (\omega_{\rm rat}( R_{\varphi^{-1}} d_1 f, d_1 g) + \omega_{\rm rat}(d_1 f, R_{\varphi^{-1}} d_1 g) ).
\end{equation*}
Comparing the Poisson bracket of the principal chiral model in this form to the Faddeev-Resheti\-khin one given by \eqref{lax PB FR}, we see that the \fr procedure \eqref{algebraic FR} equivalently reads
\beqz
(\hf, \L, R _{\varphi^{-1}}, 1) \quad \longrightarrow \quad (\hf, \L, R, 1).
\eeqz
Now it is well known \cite{Reyman:1988sf} that given an $R$-matrix $R$, for any $q \in \mathbb{C}(\!( \lambda )\!)$ the $R$-bracket associated with $R$ and $R \circ \tilde{q}$ are compatible. Therefore, taking $q = \varphi^{-1}$ in the present case we conclude that the Poisson brackets \eqref{lax PB} and \eqref{lax PB FR} are compatible.

Likewise, in the case of a symmetric space $\sigma$-model, the generalized \fr procedure can be understood as
\begin{equation*}
(\hfs, \L, R, \lambda^{-1} \phi) \; \sim \; (\hfs, \L, R_{\phi^{-1}}, \lambda^{-1}) \quad \longrightarrow \quad (\hfs, \L, R, \lambda^{-1}).
\end{equation*}
Note that in the case of a twisted loop algebra such as $\hfs$, the conclusion about compatibility of the $R$-brackets associated with $R$ and $R \circ \tilde{q}$ remains valid provided we use a $q \in \CC(\!( \lambda )\!)$ such that $q(\lambda) = q(- \lambda)$. In particular we may take $q = \phi^{-1}$, from which the desired compatibility follows.

\section{Dynamics and Hamiltonian\label{sec: Dynamics}}

Having made the choice of the Poisson bracket \eqref{FR Z2}, and following 
the method used by Faddeev and Reshetikhin as given in
\cite{Faddeev:1985qu}, we now study the dynamics of the symmetric space $\sigma$-model 
on $F/G$ where $G$ is the Lie group corresponding to $\g = \f^{(0)}$ and determine which 
reduction of the field equations may be obtained in a Hamiltonian framework. 

\subsection{Original dynamics and gauge invariance}

Let us start by recalling the expression for the Hamiltonian of the $F/G$ coset $\sigma$-model. The theory is conformally invariant at the classical level and the components of the stress-energy tensor are\footnote{We have taken the relation between the inner product and the trace of the product in some representation to be $\langle A, B\rangle= - \tr(AB) $. This leads to a positive inner product when $F$ is compact.}
\beqz
T_{\pm\pm} = - \qa \tr\bigl(A_\pm^{(1)}A_\pm^{(1)}\bigr)
\eeqz
where $A_\pm^{(1)} \eqm  \pp^{(1)} \mp A^{(1)}$. In terms of these the Hamiltonian reads
\beqz
H = \int d\sigma \bigl(  T_{++} + T_{--} +  \tr\bigl( A^{(0)} \pp^{(0)} \bigr) + \tr\bigl(\ell \, \pp^{(0)} \bigr)\bigr).
\eeqz
The field $\pp^{(0)}$ is the constraint associated with the coset gauge invariance. Its Lagrange multiplier $\ell$ takes values in $\g$. The term $\tr\bigl( A^{(0)} \pp^{(0)} \bigr)$ has been taken into account in order to be consistent with the analysis of \cite{Vicedo:2009sn} on the Hamiltonian Lax connection. By construction, the constraint $\pp^{(0)}$ is preserved by the dynamics generated by the Hamiltonian $H$. The equations of motion for the other variables $(A^{(0)}, A_\pm^{(1)})$ are, up to terms proportional to the constraint $\pp^{(0)}$,
\begin{subequations} \label{original eom}
\begin{align}
\partial_- A^{(0)} &= - \ha [A_+^{(1)}, A_-^{(1)}]
 + \partial_\sigma \ell + [ \ell, A^{(0)}] ,  \label{eqhd}   \\
 \partial_- A_+^{(1)}  &=  [ \ell, A_+^{(1)}],  \label{oap1}\\
 \partial_+ A_-^{(1)}  &= - [A_-^{(1)}, 2 A^{(0)} + \ell] \label{eqhf}
\end{align}
\end{subequations}
where $\partial_\pm = \partial_\tau \pm \partial_\sigma$. The fact 
that the dynamics depends on the arbitrary function $\ell$ of 
$\sigma$ and $\tau$ is a reflection of the gauge invariance 
generated by the constraint $\pp^{(0)}$. The corresponding 
gauge transformation of the fields reads
\begin{equation} \label{gt1}
\begin{split}
\delta A^{(0)} &=  [\alpha_R, A^{(0)}] + \partial_\sigma \alpha_R,\\
\delta A_\pm^{(1)}&=  [\alpha_R, A_\pm^{(1)}],\\
\delta \ell &= \partial_- \alpha_R  + [\alpha_R, \ell]
\end{split}
\end{equation}
with $\alpha_R$ a function taking values in $\g$. The index $R$ is used to emphasize that we are considering a right coset $F/G$. Note that a direct consequence of the equations of motion is that
\beq
\partial_{\mp} \tr\bigl[\bigl(A_\pm^{(1)} \bigr)^n\bigr] =0. \label{chir}
\eeq

\subsection{Casimirs of the new Poisson bracket and Pohlmeyer reduction}
\label{seccasimir}
Before attempting to reproduce the above dynamics in terms of the modified 
Poisson bracket \eqref{FR Z2}, we first need to 
identify the Casimirs of the latter. Indeed, since these quantities will necessarily remain constant in time with respect to \eqref{FR Z2}, we shall only be able to reproduce a reduction of the original dynamics where the same quantities have been set to constants.

The field $\pp^{(0)}$ is an obvious Casimir of \eqref{FR Z2}. Since it is a constraint of the symmetric space $\sigma$-model, it is natural to set the value of this Casimir to zero. Next, we have
\begin{alignat*}{2}
\{ A^{(0)}_{ \1}(\sigma), \A^{(1)}_{-\2}(\sigma') \}' &= - 2 [C^{(00)}_{\1\2},
\A^{(1)}_{-\2}(\sigma)] \delta_{\sigma \sigma'}, &\quad
\{ A^{(1)}_{ \1}(\sigma), \A^{(1)}_{-\2}(\sigma') \}' &= 0,\\
\{ \pp^{(0)}_{\1}(\sigma), \A^{(1)}_{-\2}(\sigma') \}' &= 0,
&\quad
\{ \pp^{(1)}_{\1}(\sigma), \A^{(1)}_{-\2}(\sigma') \}' &= 0.
\end{alignat*}
This implies that the quantities $\tr\bigl[\bigl(A_-^{(1)} \bigr)^n\bigr]$ are also Casimirs of this Poisson bracket. The existence of these other Casimirs is the first sign that one will have to perform a Pohlmeyer reduction of the coset $\sigma$-model. Equation \eqref{chir} shows that the densities $\tr\bigl[\bigl(A_-^{(1)} \bigr)^n\bigr]$ are chiral in the original model and may therefore be set to constants. However, in order to do this reduction in a consistent manner, one must take into consideration the fact that these quantities are not all independent (see for instance \cite{Evans:2000qx, Evans:2001sz}). 
We follow the references \cite{Grigoriev:2007bu, Miramontes:2008wt}. The number of independent quantities corresponds to the rank of $F/G$, which is defined to be the dimension of the maximal abelian subspaces of $\f^{(1)}$. For instance, in the case of $S^n$ or $AdS_n$ this dimension is one. This is however not the end of the story as there is a further simplification. Indeed, when $\pp^{(0)}=0$, the only non-vanishing Poisson brackets in \eqref{FR Z2} are
\begin{subequations} \label{pb A0 A-}
\begin{alignat}{2}
 \{ A^{(0)}_{ \1}(\sigma), A^{(0)}_{ \2}(\sigma') \}' &=
- 2 [C^{(00)}_{\1\2},   A^{(0)}_{\2}(\sigma) ] \delta_{\sigma \sigma'} +
2 C^{(00)}_{\1\2} \delta'_{\sigma \sigma'}, \label{pb00}\\
 \{ A^{(0)}_{ \1}(\sigma), A^{(1)}_{-\2}(\sigma') \}' &=
- 2 [C^{(00)}_{\1\2},   A^{(1)}_{-\2}(\sigma) ] \delta_{\sigma \sigma}. \label{pb01}
\end{alignat}
\end{subequations}
This implies, in particular, that $A_+^{(1)}$ is yet another Casimir. At this stage it is clear that we face exactly the same situation as in the Pohlmeyer reduction. We shall therefore fix the value of each Casimir to coincide with the value of the same quantity in the Pohlmeyer reduced coset $\sigma$-model.

Let $\a$ be a maximal abelian subspace of $\f^{(1)}$. We fix the value of the Casimir $A_+^{(1)}$ by setting
\beq
 A_+^{(1)}= \mu_+ T_+, \label{pgf}
\eeq
where $T_+ \in \a$ and $\mu_+ \in \mathbb{R}$ are constant\footnote{In the case of $AdS_n$, it is possible to choose a $T_+$ that would correspond to the vanishing of the components $T_{++}$ of the stress-energy tensor. We do not consider this case here.}. Within the scheme of the Pohlmeyer reduction, condition \eqref{pgf} is the result of two separate steps (see for instance \cite{Grigoriev:2007bu, Miramontes:2008wt}). The first one corresponds to imposing the partial gauge fixing condition $A_+^{(1)}(\sigma,\tau) = \mu_+(\sigma,\tau)T_+$. The fact that this is a valid partial gauge fixing condition follows from the polar decomposition theorem\footnote{The polar decomposition theorem is only valid when the group $G$ is compact. An extension of this theorem to the case of anti-de Sitter spaces may be found for instance in \cite{Miramontes:2008wt}.}. The second step consists in fixing the on-shell chiral function $\mu_+(\sigma,\tau)$ to a constant using a holomorphic conformal transformation, which corresponds to a partial reduction of conformal symmetry. 

The adjoint action $\text{Ad}_{T_+}$ of the element $T_+ \in \a$ defines a decomposition  of  $\f$. We shall denote it as $\f = \f^{[0]} \oplus \f^{[1]}$ where
\beqz
\mathfrak{f}^{[0]} = \text{Ker}\bigl( \text{Ad}_{T_+} \bigr) \et
\mathfrak{f}^{[1]} = \text{Im}\bigl( \text{Ad}_{T_+} \bigr),
\eeqz
which satisfy $[ \f^{[0]}, \f^{[0]}] \subset \f^{[0]}$ and $[ \f^{[0]}, \f^{[1]}] \subset \f^{[1]}$. We define $\h_R = \g^{[0]}$ and let $H_R \subset G$ be the corresponding Lie subgroup which consists of elements commuting with $T_+$. For instance, in the case of $S^n = SO(n+1)/SO(n)$ we have $H_R \simeq SO(n-1)$. We immediately see from \eqref{oap1} that the stability of condition \eqref{pgf} under time evolution requires $[T_+,  \ell] = 0$, \emph{i.e.} $\ell \in \h_R$ or equivalently $\ell = \ell^{[0]}$. Likewise, it then follows from \eqref{gt1} that the residual gauge transformations preserving \eqref{pgf} are such that $\alpha_R \in \h_R$.

Next, we pick another element $T_-\in \mathfrak{a}$. As for $T_+$, this element defines its own decomposition of $\f$ along with a Lie algebra $\mathfrak{h}_L = \g \cap \text{Ker}(Ad_{T_-})$ and its corresponding Lie subgroup $H_L \subset G$ of elements commuting with $T_-$. Sticking to our general strategy we have to fix the values of all the Casimirs $\tr\bigl[\bigl(A_-^{(1)} \bigr)^n\bigr]$. Introducing a new field $g$ taking value in $G$ through the use of the polar decomposition theorem, this may be done by setting
\beq
A_-^{(1)} =  \mu_- g^{-1} T_- g. \label{defga}
\eeq

From the point of view taken in this article, the fact that $A_-^{(1)}$ is related by the adjoint action of $g$ to a constant matrix $\mu_-T_-$ comes from the necessity to fix the Casimirs $\tr\bigl[\bigl(A_-^{(1)} \bigr)^n\bigr]$. From the point of view of the Pohlmeyer reduction, the equation (\ref{defga}) is reached by using conformal invariance, or an extension of conformal invariance in the case where the dimension of $\mathfrak{a}$ is bigger than $1$. In both cases, however, consistency of (\ref{defga}) with the dynamics is ensured by the fact that the Casimirs $\tr\bigl[\bigl(A_-^{(1)} \bigr)^n\bigr]$ are chiral densities. However, $g$ is clearly not uniquely defined by \eqref{defga}. Indeed, the transformation
\beq
g \to h_L \, g \label{invL},
\eeq
where $h_L$ is a $H_L$-valued function, leads to the same $A_-^{(1)}$.

In the following we shall restrict to the case where there exists an automorphism $\iota$ of the algebra ${\mathfrak f}$ which relates $T_-$ and $T_+$ as $T_+ = \iota(T_-)$, and with the following properties
\beq \iota({\mathfrak f}^{(i)})\subset {\mathfrak f}^{(i)},\, i=0,1 \et
\tr\left(\iota(M)\iota(N)\right)=\tr\left(MN\right),\, M,N\in{\mathfrak f}.\label{priota}
\eeq
In particular, the automorphism $\iota$ relates the right algebra ${\mathfrak h}_R$ with the left algebra ${\mathfrak h}_L$, ${\mathfrak h}_R=\iota({\mathfrak h}_L)$.

\subsection{Lifting to $G$}

\subsubsection{Poisson brackets}

At this point the phase space is parametrized by $g$ and $A^{(0)}$ taking values in $G$ and $\g$ respectively. We therefore need to lift the Poisson brackets \eqref{pb A0 A-} to the pair of fields $(A^{(0)}, g)$. To avoid clutter, in the rest of this section we suppress superscripts corresponding to the $\mathbb{Z}_2$-grading, \emph{i.e.} we write $A \equiv A^{(0)}$, $A_- \equiv A_-^{(1)}$ and
\begin{subequations} \label{pballag}
\begin{align}
 \{ A_{-\1}(\sigma), A_{-\2}(\sigma')) \}' &=0, \label{ap1ap1}\\
 \{ A_{-\1} (\sigma), A_{\2} (\sigma') \}' &= 2 [C_{\1\2}^{(00)}, A_{-\1}(\sigma) ]  \delta_{\sigma \sigma'}. \label{ap1a0}
\end{align}
\end{subequations}
The first Poisson bracket \eqref{ap1ap1} is clearly satisfied if we let $\{ g(\sigma), g(\sigma') \}' =0$. Consider now the second Poisson bracket \eqref{ap1a0}. Using the relation, following from \eqref{defga},
\beq
\{ A_{-\1}(\sigma), A_{\2}(\sigma') \}' = \bigl[A_{-\1}(\sigma) , g_{\1}(\sigma)^{-1} \{g_{\1}(\sigma), A_{\2}(\sigma') \}' \bigr], \label{pbgan0}
\eeq
we observe that \eqref{ap1a0} is satisfied if we introduce the following Poisson bracket
\beqz
\{ g_{\1}(\sigma), A_{\2}(\sigma') \}'
=  -2 g_{\1}(\sigma) C_{\1\2}^{(00)} \delta_{\sigma \sigma'}.
\eeqz
The complete set of Poisson brackets between the fields $g$ and $A$ therefore reads
\begin{subequations} \label{319}
\begin{align}
 \{ g_{\1}(\sigma), g_{\2}(\sigma') \}' &= 0, \label{cpb1}\\
\{ g_{\1}(\sigma), A_{\2}(\sigma') \}' &=  -2 g_{\1}(\sigma) C_{\1\2}^{(00)} \delta_{\sigma \sigma'},\\
\{A_{\1}(\sigma), A_{\2}(\sigma') \}' &= -2 [C_{\1\2}^{(00)}, A_{\2}(\sigma)] \delta_{\sigma \sigma'} + 2 C_{\1\2}^{(00)} \delta'_{\sigma \sigma'}. \label{cpb3}
\end{align}
\end{subequations}
This is precisely the canonical Poisson brackets of the WZW model with group $G$ (see \emph{e.g.} \cite{Bowcock:1988xr}).

\subsubsection{Reduced original dynamics}

The equations \eqref{eqhd} and \eqref{eqhf} that remain after imposing \eqref{pgf} and \eqref{defga} are
\begin{subequations} \label{322}
\begin{align}
\partial_-  A^{[0]} &= \partial_{\sigma} \ell + \bigl[ \ell , A^{[0]}\bigr] ,  \label{124}\\
\partial_- A^{[1]} &= - \ha \mu_+\mu_- \bigl[T_+, g^{-1} T_- g\bigr] +  \bigl[ \ell , A^{ [1] }\bigr], \label{125} \\
\partial_+ A_- &= - \bigl[A_- , 2 A + \ell\bigr].
\label{eqap1red}
\end{align}
\end{subequations}
Here we have extracted the components of \eqref{eqhd} along $\f^{[0]}$ and $\f^{[1]}$ using the properties of these spaces and the fact that $\ell \in \h_R$.
These equations of motion are invariant under the infinitesimal gauge transformations
\begin{subequations} \label{325}
\begin{align}
 \delta A^{[0]} &=  [\alpha_R, A^{[0]}] + \partial_\sigma \alpha_R,\label{gaga1}\\
 \delta A^{[1]} &=  [\alpha_R, A^{[1]}],\\
\delta \ell &=
\partial_- \alpha_R + [\alpha_R, \ell],\label{gaga2}\\
\delta g &= \alpha_L g - g \alpha_R \quad \Longrightarrow \quad
\delta A_-=  [\alpha_R, A_-]\label{gaga3}
\end{align}
\end{subequations}
where the functions $\alpha_R$ and $\alpha_L$ take values in $\h_R$ and $\h_L$ respectively. Next, we lift the equation of motion \eqref{eqap1red} to an equation of motion for $g$. Using the property $\partial_+ A_- = [A_-, g^{-1} \partial_+ g]$, we find
\begin{equation*}
\Bigl[T_-, \partial_+g g^{-1}  + g \bigl( 2  A  + \ell \bigr) g^{-1} \Bigr] = 0.
\end{equation*}
This shows that the equation of motion for $g$ is
\beq
 \partial_+ g g^{-1} + g \bigl(
2 A + \ell \bigr) g^{-1} = \widetilde{\ell} , \label{eqga}
\eeq
where $\widetilde{\ell}$ is an arbitrary function taking value in the subalgebra $\h_L \subset \g$.
Its presence reflects the appearance of the left gauge invariance \eqref{invL} upon introducing the phase space field $g$ to replace $A_-$.
The equation \eqref{eqga} is invariant under the complete set of gauge transformations \eqref{325} provided that the function $\widetilde{\ell}$ transforms as
\beq
\delta\widetilde{\ell} =\partial_+\alpha_L+[\alpha_L,\widetilde{\ell} ].\label{transl}
\eeq
In other words, $\widetilde{\ell}$ behaves as a gauge field for the gauge invariance under the left group $H_L$. For later use, 
let us write down the expression  of $A$ coming from equation \eqref{eqga}
\begin{equation}
A = \ha \bigl( g^{-1} \widetilde{\ell} g - g^{-1}\partial_+g - \ell \bigr).\label{At}
\end{equation}
We shall need, in particular, the projection of this equation to $\h_R$
\begin{equation}
A^{[0]} = \ha \bigl( -\ell+(g^{-1}\widetilde{\ell}g-g^{-1}\partial_+g)^{[0]} \bigr). \label{Ah}
\end{equation}

\subsubsection{Gauge invariances and anomalies}

We are now ready to look for a functional which generates the infinitesimal gauge transformations \eqref{325} on the phase space fields $g$ and $A$ in terms of the Poisson bracket \eqref{319}. This is easily found to be
\beq
\ha \int d\sigma \tr\bigl( \alpha_L J - A \alpha_R \bigr) \label{genl}
\eeq
where the current $J$ is defined by
\beq
J = \partial_\sigma g g^{-1} + g A g^{-1}. \label{curL}
\eeq
Indeed, a simple computation leads to the following Poisson brackets
\begin{align*}
 \{ J_{\1}(\sigma), g_{\2}(\sigma') \}' &= 2 C_{\1\2}^{(00)} g_{\2} (\sigma) \delta_{\sigma \sigma'},\\
\{ J_{\1}(\sigma), A_{\2} (\sigma') \}' &=0.
\end{align*}
The second equation means that gauge transformations with parameter $\alpha_L$ do not act on the field $A$, as it should be, but only generate a left multiplication on $g$. Furthermore, the current $J$ satisfies
\beq
\{ J_{\1}(\sigma), J_{\2}(\sigma') \}' = 2 [C_{\1\2}^{(00)}, J_{\2}(\sigma)] \delta_{\sigma \sigma'} - 2 C_{\1\2}^{(00)} \delta'_{\sigma \sigma'}. \label{jjpb}
\eeq
At this stage we have field equations which possess a gauge 
invariance with gauge group $H_L\times H_R$, the infinitesimal 
gauge transformations of which are respectively generated by 
the field $J$ restricted to $\h_L$ and the field $A$ restricted to 
$\h_R$. However, neither of these generators have first class 
Poisson brackets, as is apparent from \eqref{jjpb} and \eqref{cpb3}. 
This is just a reflection of the well-known fact that left and 
right isometries of the Wess-Zumino-Witten model cannot be 
freely gauged \cite{Witten:1991mm}. However, we also know that a diagonal subgroup of left and right isometries may be gauged. In our case, we indeed find that the combination $\iota(J)-A$ of the left and right generators has first class Poisson brackets. Using \eqref{jjpb} and \eqref{cpb3} along with the properties \eqref{priota} of the automorphism $\iota$, we get
\beq
\{\iota(J_{\1}(\sigma)) - A_{\1}(\sigma), \iota(J_{\2}(\sigma')) - A_{\2}(\sigma') \}' = 2 \bigl[ C_{\1\2}^{(00)}, \iota(J_{\2}(\sigma)) - A_{\2}(\sigma) \bigr] \delta_{\sigma \sigma'}.  \label{japb}
\eeq
Before looking for a Hamiltonian formulation of the field equations \eqref{124}, \eqref{125} and \eqref{eqga}, we will thus be led to partially fix the gauge invariance. In this process we shall need the following expression for the part of $\iota(J)$ lying in $\h_R$, which is easily established using the field equation \eqref{eqga} and the definition \eqref{curL},
\begin{equation}
\iota(J)^{[0]} = \ha \bigl( \iota(\widetilde\ell)-\iota(g \ell g^{-1}-\partial_-gg^{-1})^{[0]} \bigr). \label{Jh}
\end{equation}

\subsection{New Hamiltonian}

\subsubsection{Partial gauge fixing on equations of motion}

A short calculation shows that the field $J$ transforms under the gauge transformations \eqref{325} as
\beq
\delta J = \partial_\sigma\alpha_L+[\alpha_L,J]. \label{transJ}
\eeq
We are going to fix the part of the gauge invariance characterized by the relation $\iota(\alpha_L) = -\alpha_R$. Under such gauge transformations, using (\ref{gaga1}) and (\ref{transJ}), we find the transformation 
\beqz
\delta(\iota(J)-A)=-2\partial_\sigma\alpha_R-[\alpha_R,(\iota(J)+A)]
\eeqz
We take as partial gauge condition
\beq
(\iota(J)-A)^{[0]} = 0. \label{clr}
\eeq
This leaves a gauge invariance where the left and right transformations
are now related by
\beqz
\iota(\alpha_L) = \alpha_R.
\eeqz
Let us note that the equation \eqref{clr} comes out as a partial gauge fixing condition in our 
study of the field equations, and that it will also play the role of a constraint generating the remaining
gauge invariance in the Hamiltonian framework that will soon be described.
The last step before determining the Hamiltonian is to work out the consequences of the constraint \eqref{clr}. Starting from the definition \eqref{curL} of $J$ and using the equations of motion \eqref{124}, \eqref{125} and \eqref{eqga} for $A$ and $g$ we derive the equation of motion for $J$ to be
\beqz
\partial_+ J   = \partial_{\sigma} \widetilde{\ell} + \bigl[ \widetilde{\ell}, J\bigr] - \ha \mu_+ \mu_- \bigl[ g T_+ g^{-1} , T_- \bigr].
\eeqz
Combining this with the equation of motion \eqref{124} for $A^{[0]}$, we find that the constraint \eqref{clr} is preserved by the dynamics if
\beq
\iota(\widetilde{\ell})= \ell + 2 \iota(J)^{[{0}]}. \label{ltc}
\eeq
Using the equations (\ref{Ah}) and (\ref{Jh}), the gauge 
constraint (\ref{clr}) and its dynamical consequence (\ref{ltc}) 
may equivalently be written as the set of two equations
\begin{subequations} \label{field eqs}
\begin{align}
\ell &= \iota\left(g \ell g^{-1} + \partial_- g g^{-1}\right)^{[0]}, \label{fe1}\\
\iota(\widetilde{\ell}) &= \left(g^{-1}\widetilde{\ell}g - g^{-1}\partial_+g\right)^{[0]}. \label{fe2}
\end{align}
Finally, using the expression (\ref{At}) for the field $A$ enables us to rewrite the second order field equations \eqref{124} and \eqref{125} as 
\begin{equation}
\Bigl[ \partial_- - \ell, \partial_+ + g^{-1} \partial_+ g - g^{-1} \widetilde{\ell} g \Bigr] = \mu_+ \mu_- \bigl[T_+, g^{-1}T_-g\bigr]. \label{fe3}
\end{equation}
\end{subequations}
The equations of motion \eqref{field eqs} coincide with the equations of motion of a $G/H$ gauged WZW model with a potential, where the asymmetric coset $G/H$ is defined as
\beqz
G/H = G / \bigl[ g \sim h_L g \iota(h_L^{-1}) \; | \; h_L \in H_L\bigr].
\eeqz
Making the comparison for instance with \cite{Miramontes:2008wt}, we see agreement with equations (3.40)-(3.41) there provided we identify the variables ${\cal A}_{\pm}$ there as ${\cal A}_+ = -\widetilde{\ell}$ and $ {\cal A}_- = - \ell$.

\subsubsection{Hamiltonian}

At last we are ready to describe the above dynamics on the phase space parametrized by the fields $A$ and $g$ with respect to the Poisson bracket \eqref{319}, taking into account the first class constraint \eqref{clr}. It is easy to check that the Hamiltonian is given by
\begin{equation}
H' \! = \!  \int d\sigma \tr\bigl(
- \qa  \iota(J)^{[{1}]} \iota(J)^{[{1}]} - \qa A^{[1]} A^{[1]}
+ \ha \bigl(\ell  + A^{[0]} \bigr) \bigl( \iota(J)^{[{0}]} - A^{[0]} \bigr)
+ \qa \mu_+\mu_- g^{-1}T_- g T_+ \bigr).  \label{nh0}	
\end{equation}
Indeed, its Poisson brackets with $g$ and $A$ are, up to terms proportional to the constraint 
\eqref{clr},
\begin{subequations} \label{3subeq}
\begin{align}
\{ H', g \}' &= \widetilde{\ell} g - g \ell - J g - g A,\\
\{ H', A \}' &= \partial_{\sigma} A + \partial_{\sigma} \ell + [\ell, A] - \ha \mu_+ \mu_- [T_+, g^{-1} T_- g].
\end{align}
\end{subequations}
Here we have introduced $\widetilde{\ell}$ through the 
equation $\iota(\widetilde{\ell}) = \ell + 2 A^{[0]}$ which is 
equivalent to \eqref{ltc} and made use of the properties \eqref{priota} 
of the automorphism $\iota$. The first equation gives the Hamiltonian 
form of \eqref{eqga} after substituting the definition \eqref{curL} of $J$, 
and the second equation is equivalent to the Hamiltonian form of 
\eqref{124} and \eqref{125}.
Let us finally note that the Hamiltonian \eqref{nh0} is consistent with the result (3.23) of \cite{Bowcock:1988xr}. 

\subsubsection{Lax Pair}

Recall from section \ref{sec: FR bracket} that by definition the generalized Faddeev-Reshetikhin model has the same Lax matrix $\L(\lambda)$ as the original theory. In its expression \eqref{laxfg} for coset $\sigma$-models, we can of course replace the Casimirs $\Pi^{(0)}$, $A_+^{(1)}$ by their chosen values and $A_-^{(1)}$ by its expression \eqref{defga} through which $g$ is defined. This has no effect on the Poisson bracket of $\L(\lambda)$, which is therefore of the form \eqref{pb1} with $r_{\1\2}$ and $s_{\1\2}$ given explicitly later in \eqref{rs kernels}. Thus, we have\footnote{Here we restore the superscript notation corresponding to the grading $\f = \f^{(0)} \oplus \f^{(1)}$.} 
\beqz
\L(\lambda) = A^{(0)} + \ha \lambda^{-1} \mu_- g^{-1} T_- g - \ha \lambda \mu_+ T_+.
\eeqz
The expression for the temporal component $\M$ of the Lax pair similarly reads \cite{Vicedo:2009sn}
\beqz
\M(\lambda) = A^{(0)} + \ell - \ha \lambda^{-1} \mu_- g^{-1} T_- g  - \ha \lambda \mu_+ T_+.
\eeqz
The zero curvature equation $\{ H', \L \}' = \partial_\sigma \M + [\M,\L]$ is equivalent to the Hamiltonian equations of motion \eqref{3subeq}.

\medskip

The upshot of this section is that the field theory obtained from the $F/G$ coset $\sigma$-model through the generalized \fr procedure is nothing but the one corresponding to the $G/H$ gauged WZW action with an integrable potential. This also means that we have automatically obtained the $r/s$-matrices associated with the Lax matrix of these latter models. In particular, their non-ultralocality is mild.

Another key point is that even though we have made a reduction, the Poisson brackets \eqref{319} on the reduced phase space is still perfectly local. The reason being that the reduction conditions of section \ref{seccasimir} are Casimirs of the Poisson brackets \eqref{FR Z2}. This is in sharp contrast with the result of applying a similar reduction to the canonical Poisson brackets \eqref{standard Z2 brackets} of the $\sigma$-model, where the corresponding Poisson brackets on the reduced phase space turn out non-local. For the $S^2$ $\sigma$-model this has been worked out in \cite{Mikhailov:2005sy} and references therein. In the case of string theory on $AdS_5 \times S^5$, this was first studied in \cite{Mikhailov:2006uc} and then in more detail in \cite{Schmidtt:2010bi,Schmidtt:2011nr}.

\section{Towards a lattice discretization}  \label{sec: lattice}

We have seen in the previous sections that a Faddeev-Reshetikhin 
type model could also be defined for coset $\sigma$-models. The 
important novelty in this case, however, is that the modified bracket \eqref{FR Z2}
is also non-ultralocal. At first glance it might therefore seem that this 
new Poisson bracket is of no improvement compared to the original one. 
Indeed, the motivation for attempting to generalize the Faddeev-Reshetikhin 
approach to the case at hand was to try and do away with the problematic 
non-ultralocal terms occurring in the original brackets of the coset 
$\sigma$-model. Yet as we will show in the present section 
following \cite{SemenovTianShansky:1995ha}, the 
non-ultralocality of the new bracket \eqref{FR Z2} is mild compared to that 
of the original Poisson bracket of the coset $\sigma$-model. In fact,
quite remarkably, it turns out  that with this milder form of non-ultralocality
one is able to write down a corresponding well defined regularized lattice
Poisson algebra which reduces to the non-ultralocal bracket of the Lax matrix
in the continuum limit.

\subsection{Generalized Gauss decomposition}

As can be inferred from the form of the Poisson bracket \eqref{lax PB} between functions of the Lax matrix, the non-ultralocality of our model stems from the fact that its $R$-matrix, defined as in \eqref{PCM R-matrix}, is not skew-symmetric with respect to the inner product at hand \eqref{trig ip}. Nevertheless, the crucial property which will ultimately enable us to discretize the Poisson bracket of Lax matrices corresponding to \eqref{FR Z2} is that the $R$-matrix only fails to be skew-symmetric on a finite-dimensional subalgebra of the full twisted loop algebra $\hfs$ \cite{SemenovTianShansky:1995ha}. Specifically, the symmetric part of the $R$-matrix is a projection onto this subalgebra. In this case we say that the non-ultralocality of the resulting model is mild.

In order to describe the skew-symmetric and symmetric parts of $R$ with respect to \eqref{trig ip}, consider the following generalized Gauss decomposition
\begin{equation} \label{Gauss decomp}
\hfs = \hfs_{< 0} \dotplus \g \dotplus \hfs_{> 0}.
\end{equation}
Recall the definitions of the projections $\pi_{< 0}$, $\pi_0$ and $\pi_{> 0}$ from section \ref{sec: FR bracket}. In terms of these, we may write the $R$-matrix \eqref{PCM R-matrix} as $R = \pi_{> 0} + \pi_0 - \pi_{< 0}$ while its skew-symmetric and symmetric parts respectively read
\begin{equation} \label{r/s matrices}
r \eqm  \ha (R - R^{\ast}) = \pi_{> 0} - \pi_{< 0}, 
\qquad s \eqm \ha (R + R^{\ast}) = \pi_0.
\end{equation}
To see this, remember that the adjoint of the $R$-matrix with 
respect to the inner product \eqref{trig ip} is
\begin{equation} \label{R adjoint}
R^{\ast} = - \tilde{\lambda} \circ R \circ \tilde{\lambda}^{-1}.
\end{equation}
Now the subspaces $\hfs_{< 0}$, $\g$ and $\hfs_{> 0}$ respectively satisfy
\begin{equation*}
\lambda^{-1} \hfs_{< 0} \subset \hf_{< 0}, \qquad \lambda^{-1} \g 
\subset \hf_{< 0}, \qquad \lambda^{-1} \hfs_{> 0} \subset \hf_{\geq 0}.
\end{equation*}
Then decomposing any $X \in \hfs$ as $X = X_{< 0} + X_0 + X_{> 0}$ 
according to \eqref{Gauss decomp} we have
\begin{alignat*}{8}
R(\lambda^{-1} X_{> 0}) &= &&\pi_{\geq 0} (\lambda^{-1} X_{> 0}) &&= 
&&\lambda^{-1} X_{> 0} &&= &&\lambda^{-1} \pi_{> 0}(X_{> 0}) &&=
&&\lambda^{-1} R(X_{> 0}),\\
R(\lambda^{-1} X_0) &= - &&\pi_{< 0} (\lambda^{-1} X_0) &&=
- &&\lambda^{-1} X_0 &&= - &&\lambda^{-1} \pi_0(X_0) &&= - &&\lambda^{-1} R(X_0),\\
R(\lambda^{-1} X_{< 0}) &= - &&\pi_{< 0} (\lambda^{-1} X_{< 0}) &&= 
- &&\lambda^{-1} X_{< 0} &&= - &&\lambda^{-1} \pi_{< 0}(X_{< 0}) &&= 
&&\lambda^{-1} R(X_{< 0}).
\end{alignat*}
Now combining this result with \eqref{R adjoint} we see that the restriction of $R$ to the subspace $\hfs_{< 0} \dotplus \hfs_{> 0}$ is skew-symmetric whereas its restriction to $\g$ is symmetric, from which \eqref{r/s matrices} follows.

Now let $\alpha \in \text{End}\, \g$ be any skew-symmetric solution\ of mCYBE \eqref{mybm} on $\g$, with $\omega=1$. In particular $-\alpha$ is also a solution. It is straightforward to check by a direct calculation that the operators
\begin{equation*}
r \pm \alpha = \text{diag}(-1, \pm \alpha, 1) \in \text{End}\, \hfs
\end{equation*}
where the diagonal decomposition is relative to \eqref{Gauss decomp}, 
are both skew-symmetric solutions of mCYBE on $\hfs$. In other words 
we have
\begin{subequations} \label{mCYBE on double}
\begin{equation} \label{mCYBE on double a}
[(r \pm \alpha) X, (r \pm \alpha) Y] - (r \pm \alpha) \bigl( [(r \pm \alpha) X, Y] 
+ [X, (r \pm \alpha) Y] \bigr) + [X, Y] = 0,
\end{equation}
for any $X, Y \in \hfs$. 
It is important to stress that $r$ itself is \emph{not} a solution of mCYBE. Furthermore one can also easily check that the matrices $s \pm \alpha$ satisfy the following relations with $r \pm \alpha$,
\begin{equation} \label{mCYBE on double b}
[(s \pm \alpha) X, (s \pm \alpha) Y] = (s \pm \alpha) \bigl( [(r \pm \alpha) X, Y] 
+ [X, (r \pm \alpha) Y] \bigr).
\end{equation}
\end{subequations}
Again we stress that if we set $\alpha = 0$ these relations no longer hold.

One can write down explicit kernels for the operators $r \pm \alpha$ 
and $s \pm \alpha$ as follows. The kernels for the projection operators 
$\pi_{<0}$, $\pi_0$ and $\pi_{>0}$ respectively read
\begin{equation*}
\pi^{<0}_{\1\2} (\lambda, \mu) = \sum_{m = 1}^{\infty} \left( \frac{\mu}{\lambda} 
\right)^m C_{\1\2}^{(m \, m)}, \quad
\pi^0_{\1\2}(\lambda, \mu) = C_{\1\2}^{(00)}, \quad
\pi^{>0}_{\1\2} (\lambda, \mu) = \sum_{m = 1}^{\infty} \left( \frac{\lambda}{\mu} 
\right)^m C_{\1\2}^{(m \, m)}.
\end{equation*}
It then follows that the kernels of the $r/s$ matices in \eqref{r/s matrices} 
are \cite{Sevostyanov:1995hd}
\begin{equation} \label{rs kernels}
r_{\1\2} (\lambda, \mu) = \frac{\mu^2 + \lambda^2}{\mu^2 - \lambda^2} 
C_{\1\2}^{(00)} + \frac{2 \lambda \mu}{\mu^2 - \lambda^2} C_{\1\2}^{(11)}, 
\qquad
s_{\1\2}(\lambda, \mu) = C_{\1\2}^{(00)}.
\end{equation}

\subsection{Lattice algebra}

A standard way of constructing an integrable lattice discretization 
of a field theory on the circle is as follows. 
Recall that the zero curvature equation arises as the compatibility condition of the following auxiliary linear system
\begin{equation} \label{aux lin sys 2}
\partial_{\sigma} \psi = \L \psi, \qquad \partial_{\tau} \psi = \M \psi.
\end{equation}
To discretize the spacial direction we replace the first equation by 
its discrete counterpart. This means breaking up the circle at a finite 
set of points $\sigma_n \in S^1$, $n = 1, \ldots, N$ and considering 
the value of $\psi$ only at these points by defining $\psi_n \coloneqq 
\psi(\sigma_n) \in \hFs$, where $\hFs$ is the loop group corresponding 
to $\hfs$. The lattice Lax matrix $\L^n$ is then defined to be the 
parallel transporter from the site $\sigma_n$ to the next site 
$\sigma_{n+1}$, namely
\begin{equation*}
\L^n = P \overleftarrow{\exp} \int_{\sigma_n}^{\sigma_{n+1}} \L(\sigma) d\sigma.
\end{equation*}
The spacial discretization of the auxiliary linear system \eqref{aux lin sys 2} then takes the following form
\begin{equation} \label{aux lin sys lattice}
\psi_{n+1} = \L^n \psi_n, \qquad \partial_{\tau} \psi_n = \M^n \psi_n,
\end{equation}
where the second equation is obtained by evaluating the last equation of \eqref{aux lin sys 2} at $\sigma_n$. We note here that $\L^n$ takes value in $\hFs$ whereas $\M^n$ still takes value in $\hfs$.

An important object in the continuum theory is the so called monodromy matrix $T$, defined as the parallel transporter around the full circle. It can be recovered on the lattice by multiplying all the lattice Lax matrices as
\begin{equation} \label{monodromy lattice}
T = \L^N \ldots \L^1.
\end{equation}
The importance of this object stems from the fact that its spectral invariants $\tr (T^p) $ generate integrals of motion of the continuum theory.

Having defined the lattice Lax matrices $\L^n$, the next step would 
be to determine their pairwise Poisson brackets. Unfortunately, 
recall from section \ref{sec: NUL prob} that when dealing with a 
non-ultralocal theory, the presence of $\delta'_{\sigma \sigma'}$ 
terms in the Poisson algebra of the continuum Lax matrix $\L(\sigma)$ 
prevents us from computing this directly. We therefore ask the reverse 
question, namely: does there exist a Poisson bracket $\{ \L_{\1}^n, \L_{\2}^m \}'$ 
satisfying all the necessary properties? First of all, this Poisson bracket 
should certainly be anti-symmetric and satisfy the Jacobi identity, \emph{i.e.}
\begin{equation*}
\{ \L_{\1}^n, \L_{\2}^m \}' = - \{ \L_{\2}^m, \L_{\1}^n \}' \et
\{ \L_{\1}^m, \{ \L_{\2}^n, \L_{\3}^p \}' \}' + \text{cycl.} = 0,
\end{equation*}
for $m,n,p = 1, \ldots, N$.
Secondly, after requiring the Leibniz rule to hold, the resulting Poisson 
bracket of the monodromy matrix $T$ should be such that the 
integrals of motion are in involution. This means that
\begin{equation} \label{involution}
\{ \tr (T^p), \tr (T^q) \}' = 0,
\end{equation}
for any positive integers $p$ and $q$.
The general quadratic Poisson algebra satisfying these requirements has been identified in \cite{Freidel:1991jx, Freidel:1991jv}.
However, the Poisson bracket between $\L^n$ and $\L^m$ should also reduce 
to the original Poisson algebra \eqref{pb1} in the continuum limit.
Remarkably, it turns out that such a Poisson bracket does exist in the 
present case \cite{SemenovTianShansky:1994dm, SemenovTianShansky:1995ha}. It can be defined in 
terms of the matrices $r \pm \alpha$ and $s \pm \alpha$ as follows
\begin{subequations} \label{lattice algebra}
\begin{align}
\label{lattice algebra a}
\{ \L^n_{\1}, \L^n_{\2} \}' &= (r + \alpha)_{\1\2} \L^n_{\1} \L^n_{\2} - \L^n_{\1} 
\L^n_{\2} (r - \alpha)_{\1\2},\\
\label{lattice algebra b}
\{ \L^n_{\1}, \L^{n+1}_{\2} \}' &= \L^{n+1}_{\2} (s - \alpha)_{\1\2} \L^n_{\1},\\
\label{lattice algebra d}
\{ \L^n_{\1}, \L^m_{\2} \}' &= 0, \qquad |n - m| \geq 2.
\end{align}
\end{subequations}
It has been shown in \cite{SemenovTianShansky:1995ha, SemenovTianShansky:1994dm} that
\eqref{lattice algebra} is the unique algebra satisfying the above requirements.
We refer the reader to \cite{SemenovTianShansky:1995ha, SemenovTianShansky:1994dm} for details and will
content ourselves here with showing that the algebra \eqref{lattice algebra}
does indeed satisfy all the desired properties.
First of all, we note that this is a well defined algebra \cite{Freidel:1991jx, Freidel:1991jv}.
In particular, anti-symmetry follows 
using the (skew-)symmetry properties of $r, s$ and $\alpha$, and it 
satisfies the Jacobi identity by virtue of the relations \eqref{mCYBE on double}
\cite{Freidel:1991jx, Freidel:1991jv}. 
Next, using the Leibniz rule to compute the Poisson bracket of two 
monodromy matrices \eqref{monodromy lattice} we find
\begin{equation} \label{monodromy algebra}
\{ T_{\1}, T_{\2} \}' = (r + \alpha)_{\1\2} T_{\1} T_{\2} - T_{\1} T_{\2} (r - \alpha)_{\1\2} - T_{\1} (s + \alpha)_{\1\2} T_{\2} + T_{\2} (s - \alpha)_{\1\2} T_{\1}.
\end{equation}
In deriving this relation, to cancel many terms we make essential 
use of the trivial but important fact that $(r + \alpha)_{\1\2} + 
(s - \alpha)_{\1\2} = (s + \alpha)_{\1\2} + (r - \alpha)_{\1\2}$. Recall also that 
we are assuming periodic boundary conditions. It is now 
easy to deduce from \eqref{monodromy algebra} that the involution 
property \eqref{involution} holds.

Finally, we must show that \eqref{lattice algebra} reduces to the 
correct continuum algebra of Lax matrices we started with when 
the lattice spacing goes to zero \cite{SemenovTianShansky:1994dm, SemenovTianShansky:1995ha}. For this we note that the lattice 
algebra \eqref{lattice algebra} can equivalently be written as a single equation
\begin{align} \label{lattice algebra 2}
\{ \L^n_{\1}, \L^m_{\2} \}' &=
(r + \alpha)_{\1\2} \L^n_{\1} \L^m_{\2} \delta_{m n}
- \L^n_{\1} \L^m_{\2} (r - \alpha)_{\1\2} \delta_{m n} \notag\\
&\qquad\qquad - \L^n_{\1} (s + \alpha)_{\1\2} \L^m_{\2} \delta_{m+1, n}
+ \L^m_{\2} (s - \alpha)_{\1\2} \L^n_{\1} \delta_{m, n+1}.
\end{align}
In order to take the continuum limit of this equation we write $\L^n = 
{\bf 1} + \Delta \L(\sigma_n) + O(\Delta^2)$ where $\Delta = 
\sigma_{n+1} - \sigma_n$ is the lattice spacing. 
Substituting this into \eqref{lattice algebra 2} and working to 
lowest order in $\Delta$ gives
\begin{align*}
\{ \L_{\1}(\sigma_n), \L_{\2}(\sigma_m) \}' &= \Delta^{-1} [r_{\1\2}, 
\L_{\1}(\sigma_n) + \L_{\2}(\sigma_m)] \delta_{m n}\\
&+ \Delta^{-1} \L_{\2}(\sigma_m) s_{\1\2} \delta_{m, n+1}
+ \Delta^{-1} s_{\1\2} \L_{\1}(\sigma_n) \delta_{m, n+1}\\
&- \Delta^{-1} \L_{\1}(\sigma_n) s_{\1\2} \delta_{m+1, n}
- \Delta^{-1} s_{\1\2} \L_{\2}(\sigma_m)  \delta_{m+1, n}\\
&- \Delta^{-1} \alpha_{\1\2} \L_{\2}(\sigma_m) (\delta_{m+1, n} - 
\delta_{m n}) + \Delta^{-1} \L_{\2}(\sigma_m) \alpha_{\1\2} (\delta_{m n} - 
\delta_{m, n+1})\\
&+ \Delta^{-1} \alpha_{\1\2} \L_{\1}(\sigma_n) (\delta_{m n} - \delta_{m, n+1}) 
- \Delta^{-1} \L_{\1}(\sigma_n) \alpha_{\1\2} (\delta_{m+1, n} - \delta_{m n})\\
&- \Delta^{-2} \alpha_{\1\2} (\delta_{m, n+1} - 2 \delta_{m n} + \delta_{m+1, n}) 
- \Delta^{-2} s_{\1\2} (\delta_{m+1, n} - \delta_{m, n+1}).
\end{align*}
In the continuum limit $\Delta \to 0$ we let $\sigma_n = \sigma$, $\sigma_m = \sigma'$ and make use of the following identities
\begin{align*}
\Delta^{-1} \delta_{mn} &\to \delta_{\sigma \sigma'}, \\
\Delta^{-1} (\delta_{m+1, n} - \delta_{mn}) &\sim \Delta 
\delta'_{\sigma \sigma'} \to 0, \\
\Delta^{-2} (\delta_{m+1, n} - 2 \delta_{mn} + \delta_{m, n+1}) 
&\sim \Delta \delta''_{\sigma \sigma'} \to 0, \\
\Delta^{-2} (\delta_{m+1, n} - \delta_{m, n+1}) &\to - 2 \delta'_{\sigma \sigma'}.
\end{align*}
Note also that all the $O(1)$ terms in the above algebra vanish in this limit since they are multiplied by some $\delta_{mn}$ which effectively goes like $\Delta \delta_{\sigma \sigma'}$. Taking $\Delta \to 0$ we therefore arrive at the following continuum algebra
\begin{equation*}
\{ \L_{\1}(\sigma), \L_{\2}(\sigma') \}' = [r_{\1\2}, \L_{\1}(\sigma) 
+ \L_{\2}(\sigma)] \delta_{\sigma \sigma'} + [s_{\1\2}, \L_{\1}(\sigma) 
- \L_{\2}(\sigma)] \delta_{\sigma \sigma'} + 2 s_{\1\2} \delta'_{\sigma \sigma'}
\end{equation*}
which is precisely the Poisson algebra \eqref{pb1} of the Lax matrix $\L(\sigma)$. In particular, we notice that all dependence on the matrix $\alpha$ disappears in the continuum. Nevertheless, without $\alpha$ the lattice algebra we wrote down does not correspond to a well defined Poisson bracket of the $\L^n$'s.

For consistency we should also check that the Poisson bracket of two parallel transporters with distinct end-points agrees with its direct computation since the latter is unambiguous \cite{Maillet:1985ek}. Specifically, consider the product of lattice Lax matrices on successive sites 
\beqz
T^{I,J} = \L^I\L^{I-1}\cdots\L^{J+1}\L^{J},
\eeqz
for any $I > J$. This object is merely the parallel transporter from $\sigma_J$ to $\sigma_{I+1}$. The Poisson bracket
\beqz
\{ T_{\1}^{I,J}, T_{\2}^{K,L} \}'
\eeqz
when all four $I$, $J$, $K$, $L$ are distinct may then be computed in two different ways: either using the lattice algebra \eqref{lattice algebra} or by a direct computation. One may check that the result of the lattice calculation is independent of $\alpha$ and moreover it agrees with the result of the direct computation. This therefore shows that the matrix $\alpha$ only enters the Poisson brackets of parallel transporters which would otherwise be ill-defined.

The lattice algebra \eqref{lattice algebra} is to be compared to its 
ultralocal counterpart \eqref{lpb1}. From the way the $s$-matrix 
appears in \eqref{lattice algebra b} it is clear that non-ultralocality manifests itself on the lattice by 
the fact that neighbouring lattice Lax matrices no longer Poisson 
commute. Note that the Poisson algebra \eqref{lattice algebra} is 
precisely of the general  quadratic $abcd$-type discussed in 
\cite{Freidel:1991jx,Freidel:1991jv} if we let
\begin{equation*}
a_{\1\2} = (r + \alpha)_{\1\2}, \qquad b_{\1\2} = (- s - \alpha)_{\1\2}, 
\qquad c_{\1\2} = (- s + \alpha)_{\1\2}, \qquad d_{\1\2} = (r - \alpha)_{\1\2}.
\end{equation*}

\section{Comments and conclusion}
\label{sec conclu}

In this article we have generalized the first steps of the \fr procedure to symmetric space $\sigma$-models. Many comments come to mind.

\medskip

To begin with let us go back to the case of the principal chiral model on a Lie group $G$. As we showed in section \ref{se122}, for this model it is possible to completely rid the Poisson brackets of their non-ultralocality. The next step would then be to determine the new Hamiltonian. Following \cite{Faddeev:1985qu} this can be achieved by defining two functions $P_S$ and $P_T$ that act, with respect to the ultralocal Poisson brackets, as spatial derivative on $S = j_0 + j_1$ and $T = j_0 - j_1$ respectively. Although explicit expressions for $P_{S,T}$ can be obtained locally in terms of Darboux coordinates, these are less important at the quantum level where the operators $P_{S,T}$ are replaced by corresponding shift operators on the lattice. It might therefore be possible to also generalize the \fr procedure to a generic principal chiral model, even without having explicit classical expressions for the operators $P_{S,T}$. 

It is however possible to proceed differently by treating the principal chiral model on $G$ as a symmetric space $\sigma$-model on $G\times G /G_{\rm diag}$, where $G_{\rm diag}$ denotes the diagonal subgroup. The work presented here may then be applied directly to this case, yielding the $G/U(1)^r$ gauged WZW model with a potential where $r$ is the rank of $\g$. These models are known as homogenous sine-Gordon models \cite{FernandezPousa:1996hi, FernandezPousa:1997zb, Miramontes:1999hx, Dorey:2004qc, CastroAlvaredo:2001ih}. In particular, the case of the principal chiral model on $SU(2)$ corresponds to the $SU(2)/U(1)$ gauged WZW model with a potential, which upon gauge fixing the $U(1)$ invariance gives the complex sine-Gordon theory \cite{Pohlmeyer:1975nb, Lund:1976ze, Lund:1977dt, Getmanov:1977hk}. An immediate drawback of this approach to treating the principal chiral model is that one departs from the analysis of \cite{Faddeev:1985qu} since the non-ultralocality is not completely removed. However, one advantage of proceeding in this way is that contrary to the case above, the action of these theories is explicitely Lorentz invariant. Furthermore, this puts the principal chiral model and the symmetric space $\sigma$-model on the same footing since the alleviation of non-ultralocality corresponds in both cases to a Pohlmeyer reduction. 

\medskip

Of course, the equivalence between the generalized \fr model defined in this article and the original symmetric space $\sigma$-model is restricted for the moment to the classical level. Any statement about the possible fate of this equivalence at the quantum level would be premature. In fact, this issue is already rather delicate for the lattice magnetic model defined in \cite{Faddeev:1985qu}. The quantization of this model by means of the Bethe ansatz describes excitations over the reference state, whereas the physical ground state is obtained by filling in the Dirac sea of Bethe roots. The claim made in \cite{Faddeev:1985qu} is that taking the classical limit around this physical ground state reproduces the non-ultralocal Poisson structure of the $SU(2)$ principal chiral model. We refer the reader to the original article \cite{Faddeev:1985qu} as well as \cite{Destri:1989} and the  more recent article \cite{Caetano:2010zd} for tests of this claim.
To proceed along the lines of \cite{Faddeev:1985qu} in the present case, the next challenge will be to explicitly construct a lattice model. The first step in this direction consists in writing the quantum lattice algebra corresponding to \eqref{monodromy algebra} which should be a quadratic algebra of the type discussed in \cite{Freidel:1991jx, Freidel:1991jv}. In this context, it would be desirable to investigate the connection with the so called lattice WZW model (see the set of lectures \cite{Falceto:1992bf}) and Kac-Moody algebra introduced in \cite{Alekseev:1991wq, Alekseev:1992wn}.

\medskip

Independently of whether the program of generalizing the Faddeev-Reshetikhin procedure can be brought to its completion, an important byproduct of our work concerns the non-ultralocality of generalized sine-Gordon models. Although we have shown it in an indirect way, a prominent result of this article is that the non-ultralocality of such models, viewed as gauged WZW models with an integrable potential, is mild. To illustrate the significance of this result, let us focus on the complex sine-Gordon model as an example. When viewed as a gauged $SU(2)/U(1)$ WZW model plus a potential, the non-ultralocality of this model is mild. However, if we gauge fix the $U(1)$ invariance then we obtain the complex sine-Gordon action
\beqz
\int d\sigma d\tau \, \ha \, \Bigl( \frac{|\partial_\mu \psi |^2
}{1 - g^2 |\psi|^2 } - m^2 | \psi |^2 \Bigr).
\eeqz
For this action, the situation is completely different since the non-ultralocality of the corresponding Poisson structure is no longer mild and the associated $r$ and $s$ matrices are in fact dynamical \cite{Maillet:1985ek}! This suggests that it may be preferable to try to discretize these theories at the level of the gauged WZW model with an integrable potential. This is reminiscent of the study in \cite{Dorey:1994mg, Hoare:2010fb} of the $S$-matrix of the complex sine-Gordon model, where the quantum counterterm added in \cite{deVega:1981ka, deVega:1982sh, Maillet:1981pj} at one loop in order to maintain factorized scattering has a natural gauged WZW origin.

\medskip

One important motivation for the present work is of course related to the AdS/CFT correspondence \cite{Maldacena:1997re, Gubser:1998bc, Witten:1998qj} between superstring theory on $AdS_5 \times S^5$ and ${\cal N} = 4$ superconformal Yang-Mills theory (see \cite{Beisert:2010jr} for a review). Since the non-ultralocality of the superstring on $AdS_5 \times S^5$ is a major obstacle to quantizing this theory, it would certainly be very appealing if the Faddeev-Reshetikhin procedure could generalize to this context as well. In view of this one should start by extending the analysis presented here to semi-symmetric space $\sigma$-models. In fact, it has already been shown in \cite{Vicedo:2010qd} that the $r/s$ structure of the superstring on $AdS_5 \times S^5$ uncovered in \cite{Magro:2008dv} has an algebraic origin which fits precisely into the $R$-matrix approach. This is exactly the right framework to proceed along the lines presented here. The analogue of the Poisson brackets \eqref{FR Z2} in this case is under investigation.

\paragraph{Acknowledgements}

We would like to thank K. Gawedzki and J. M. Maillet for useful discussions. B.V. is extremely grateful to M. Semenov-Tian-Shansky for many valuable discussions. 
B.V. is supported by UK EPSRC grant EP/H000054/1.

\appendix

\section{Notations}

In this appendix we define some notations which are used throughout the text.

Given an operator $O$ acting on $\hf$, its kernel $O_{\1\2}$ relative to the twisted inner product \eqref{twisted ip} is defined by 
\begin{equation*}
(OX,Y)_\varphi  = (O_{\1\2} , Y \otimes X)_\varphi, \qquad \forall X, Y \in \hf.
\end{equation*}
The kernel $O^\ast_{\1\2}$ of the adjoint operator $O^\ast$ is then simply given by $O_{\2\1}$. 

When writing Poisson brackets in tensor notation we make use of the tensor Casimir $C_{\1\2}$. It can be defined as the kernel of the identity operator $\text{id} \in \text{End}\, \f$ with respect to the inner product $\langle \cdot, \cdot \rangle$ on $\f$. In other words it is defined by the following property
\begin{equation*}
\langle C_{\1\2}, x_{\2} \rangle_\2 = x_{\1},
\end{equation*}
for any $x \in \f$. It is easy to check that it satisfies the property $[C_{\1\2}, x_\1 + x_\2] = 0$. The corresponding property for any group element $g \in F$ reads
\begin{equation*}
g_\1 g_\2 C_{\1\2} =   C_{\1\2} g_\1 g_\2. 
\end{equation*}

When $\f$ is equipped with an involution $\sigma : \f \to \f$ such that $\sigma^2 = \text{id}$, this induces a direct sum decomposition $\f = \f^{(0)} \oplus \f^{(1)}$ into eigenspaces of $\sigma$. By the homomorphism property of $\sigma$, this $\mathbb{Z}_2$-grading has the property that
\begin{equation*}
[\f^{(i)}, \f^{(j)}] \subset \f^{(i+j)}.
\end{equation*}
We shall always assume that the inner product $\langle \cdot, \cdot \rangle$ on $\f$ 
respects the grading, in the sense that $\langle x^{(0)}, y^{(1)} \rangle = 0$ for any $x^{(0)} \in \f^{(0)}$ and $y^{(1)} \in \f^{(1)}$. In this case the tensor Casimir can be decomposed as $C_{\1\2} = C_{\1\2}^{(00)} + C_{\1\2}^{(11)}$ where $C_{\1\2}^{(ii)} \in \f^{(i)} \otimes  \f^{(i)}$.

\providecommand{\href}[2]{#2}\begingroup\raggedright\endgroup


\begin{thebibliography}{10}

\bibitem{faddtakh1979-1}
L.~Faddeev and L.~Takhtajan, {\it {The quantum method of the inverse problem
  and the Heisenberg XYZ-model}},  {\em Russ. Math. Surveys} {\bf 34:5} (1979)
  1168.

\bibitem{Kulish:1979if}
P.~Kulish and E.~Sklyanin, {\it {Quantum inverse scattering method and the
  Heisenberg ferromagnet}},  {\em Phys. Lett.} {\bf A70} (1979) 461--463.

\bibitem{faddsklytak1980tmp1}
L.~Faddeev, E.~Sklyanin, and L.~Takhtajan, {\it {Quantum inverse problem
  method: I}},  {\em Theor. Math. Phys.} {\bf 57} (1980) 688--706.

\bibitem{Maillet:1985fn}
J.~M. Maillet, {\it {Kac-Moody algebra and extended Yang-Baxter relations in
  the $O(N)$ non-linear sigma model}},  {\em Phys. Lett.} {\bf B162} (1985)
  137.

\bibitem{Maillet:1985ek}
J.~M. Maillet, {\it {New integrable canonical structures in two-dimensional
  models}},  {\em Nucl. Phys.} {\bf B269} (1986) 54.

\bibitem{Faddeev:1985qu}
L.~Faddeev and N.~Reshetikhin, {\it {Integrability of the principal chiral
  field model in (1+1)-dimension}},  {\em Annals of Physics} {\bf 167} (1986)
  227.

\bibitem{SemenovTianShansky:1995ha}
M.~Semenov-Tian-Shansky and A.~Sevostyanov, {\it {Classical and quantum
  nonultralocal systems on the lattice}},
  \href{http://xxx.lanl.gov/abs/hep-th/9509029}{{\tt hep-th/9509029}}.

\bibitem{Pohlmeyer:1975nb}
K.~Pohlmeyer, {\it {Integrable hamiltonian systems and interactions through
  quadratic constraints}},  {\em Commun. Math. Phys.} {\bf 46} (1976) 207--221.

\bibitem{Pohlmeyer:1979ch}
K.~Pohlmeyer and K.-H. Rehren, {\it {Reduction of the two-dimensional O(n)
  nonlinear sigma model}},  {\em J. Math. Phys.} {\bf 20} (1979) 2628.

\bibitem{Eichenherr:1979yw}
H.~Eichenherr and K.~Pohlmeyer, {\it {Lax pairs for certain generalizations of
  the sine-gordon equation}},  {\em Phys. Lett.} {\bf B89} (1979) 76.

\bibitem{Eichenherr:1979mx}
H.~Eichenherr, {\it {Infinitely many conserved local charges for the $CP^{N-1}$
  models}},  {\em Phys. Lett.} {\bf B90} (1980) 121.

\bibitem{Eichenherr:1979uk}
H.~Eichenherr and J.~Honerkamp, {\it {Reduction of the $CP^n$ nonlinear sigma
  model}},  {\em J. Math. Phys.} {\bf 22} (1981) 374.

\bibitem{D'Auria:1979tb}
R.~D'Auria, T.~Regge, and S.~Sciuto, {\it {A general scheme for bidimensional
  models with associate linear set}},  {\em Phys. Lett.} {\bf B89} (1980) 363.

\bibitem{D'Auria:1980cx}
R.~D'Auria, T.~Regge, and S.~Sciuto, {\it {Group theoretical construction of
  two-dimensional models with infinite set of conservation laws}},  {\em Nucl.
  Phys.} {\bf B171} (1980) 167.

\bibitem{D'Auria:1980xs}
R.~D'Auria and S.~Sciuto, {\it {Group theoretical construction of
  two-dimensional supersymmetric models}},  {\em Nucl. Phys.} {\bf B171} (1980)
  189.

\bibitem{Zakharov:1973pp}
V.~Zakharov and A.~Mikhailov, {\it {Relativistically invariant two-dimensional
  models in field theory integrable by the inverse problem technique.}},  {\em
  Sov. Phys. JETP} {\bf 47} (1978) 1017--1027.

\bibitem{Bakas:1995bm}
I.~Bakas, Q.-H. Park, and H.-J. Shin, {\it {Lagrangian formulation of symmetric
  space sine-Gordon models}},  {\em Phys. Lett.} {\bf B372} (1996) 45--52,
  [\href{http://xxx.lanl.gov/abs/hep-th/9512030}{{\tt hep-th/9512030}}].

\bibitem{Grigoriev:2007bu}
M.~Grigoriev and A.~A. Tseytlin, {\it {Pohlmeyer reduction of AdS$_5$ $\times$
  S$^5$ superstring sigma model}},  {\em Nucl. Phys.} {\bf B800} (2008)
  450--501, [\href{http://xxx.lanl.gov/abs/0711.0155}{{\tt arXiv:0711.0155}}].

\bibitem{Miramontes:2008wt}
J.~L. Miramontes, {\it {Pohlmeyer reduction revisited}},  {\em JHEP} {\bf 10}
  (2008) 087, [\href{http://xxx.lanl.gov/abs/0808.3365}{{\tt
  arXiv:0808.3365}}].

\bibitem{Bowcock:1988xr}
P.~Bowcock, {\it {Canonical quantization of the gauged Wess-Zumino model}},
  {\em Nucl. Phys.} {\bf B316} (1989) 80.

\bibitem{SemenovTianShansky:1983ik}
M.~Semenov-Tian-Shansky, {\it {What is a classical $r$-matrix?}},  {\em Funct.
  Anal. Appl.} {\bf 17} (1983) 259--272.

\bibitem{semenov_2002}
M.~Semenov-Tian-Shansky, {\it {Integrable systems and factorization problems}},
   \href{http://xxx.lanl.gov/abs/nlin/0209057}{{\tt nlin/0209057}}.

\bibitem{Semevov2008a}
M.~Semenov-Tian-Shansky, {\it {Integrable sytsems: the $r$-matrix approach}},
  December 2008, RIMS-1650.

\bibitem{Vicedo:2010qd}
B.~Vicedo, {\it {The classical R-matrix of AdS/CFT and its Lie dialgebra
  structure}},  {\em Lett. Math. Phys.} {\bf 95} (2011) 249--274,
  [\href{http://xxx.lanl.gov/abs/1003.1192}{{\tt arXiv:1003.1192}}].

\bibitem{Vicedo:2009sn}
B.~Vicedo, {\it {Hamiltonian dynamics and the hidden symmetries of the $AdS_5
  \times S^5$ superstring}},  {\em JHEP} {\bf 1001} (2010) 102,
  [\href{http://xxx.lanl.gov/abs/0910.0221}{{\tt arXiv:0910.0221}}].

\bibitem{Sevostyanov:1995hd}
A.~Sevostyanov, {\it {The Classical $R$ matrix method for nonlinear sigma
  model}},  {\em Int. J. Mod. Phys.} {\bf A11} (1996) 4241--4254,
  [\href{http://xxx.lanl.gov/abs/hep-th/9509030}{{\tt hep-th/9509030}}].

\bibitem{Reyman:1988sf}
A.~Reyman and M.~Semenov-Tian-Shansky, {\it {Compatible Poisson structures for
  Lax equations: an R matrix approach}},  {\em Phys. Lett.} {\bf A130} (1988)
  456--460.

\bibitem{Evans:2000qx}
J.~Evans and A.~Mountain, {\it {Commuting charges and symmetric spaces}},  {\em
  Phys. Lett.} {\bf B483} (2000) 290--298,
  [\href{http://xxx.lanl.gov/abs/hep-th/0003264}{{\tt hep-th/0003264}}].

\bibitem{Evans:2001sz}
J.~M. Evans, {\it {Integrable sigma models and Drinfeld-Sokolov hierarchies}},
  {\em Nucl. Phys.} {\bf B608} (2001) 591--609,
  [\href{http://xxx.lanl.gov/abs/hep-th/0101231}{{\tt hep-th/0101231}}].

\bibitem{Witten:1991mm}
E.~Witten, {\it {On Holomorphic factorization of WZW and coset models}},  {\em
  Commun. Math. Phys.} {\bf 144} (1992) 189--212.

\bibitem{Mikhailov:2005sy}
A.~Mikhailov, {\it {A nonlocal Poisson bracket of the sine-Gordon model}},
  {\em J. Geom. Phys.} {\bf 61} (2011) 85--94,
  [\href{http://xxx.lanl.gov/abs/hep-th/0511069}{{\tt hep-th/0511069}}].

\bibitem{Mikhailov:2006uc}
A.~Mikhailov, {\it {Bihamiltonian structure of the classical superstring in
  $AdS_5 \times S^5$}},  \href{http://xxx.lanl.gov/abs/hep-th/0609108}{{\tt
  hep-th/0609108}}.

\bibitem{Schmidtt:2010bi}
D.~M. Schmidtt, {\it {Supersymmetry Flows, Semi-Symmetric Space Sine-Gordon
  Models And The Pohlmeyer Reduction}},  {\em JHEP} {\bf 1103} (2011) 021,
  [\href{http://xxx.lanl.gov/abs/1012.4713}{{\tt arXiv:1012.4713}}].

\bibitem{Schmidtt:2011nr}
D.~M. Schmidtt, {\it {Integrability vs Supersymmetry: Poisson Structures of The
  Pohlmeyer Reduction}},  {\em JHEP} {\bf 1111} (2011) 067,
  [\href{http://xxx.lanl.gov/abs/1106.4796}{{\tt arXiv:1106.4796}}].

\bibitem{SemenovTianShansky:1994dm}
M.~Semenov-Tian-Shansky, {\it {Monodromy map and classical $R$-matrices}},
  {\em Journal of Math. Sciences} {\bf 77} (1995) 3226,
  [\href{http://xxx.lanl.gov/abs/hep-th/9402054}{{\tt hep-th/9402054}}].

\bibitem{Freidel:1991jx}
L.~Freidel and J.~M. Maillet, {\it {Quadratic algebras and integrable
  systems}},  {\em Phys. Lett.} {\bf B262} (1991) 278--284.

\bibitem{Freidel:1991jv}
L.~Freidel and J.~M. Maillet, {\it {On classical and quantum integrable field
  theories associated to Kac-Moody current algebras}},  {\em Phys. Lett.} {\bf
  B263} (1991) 403--410.

\bibitem{FernandezPousa:1996hi}
C.~R. Fernandez-Pousa, M.~V. Gallas, T.~J. Hollowood, and J.~Miramontes, {\it
  {The Symmetric space and homogeneous sine-Gordon theories}},  {\em Nucl.
  Phys.} {\bf B484} (1997) 609--630,
  [\href{http://xxx.lanl.gov/abs/hep-th/9606032}{{\tt hep-th/9606032}}].

\bibitem{FernandezPousa:1997zb}
C.~R. Fernandez-Pousa, M.~V. Gallas, T.~J. Hollowood, and J.~Miramontes, {\it
  {Solitonic integrable perturbations of parafermionic theories}},  {\em Nucl.
  Phys.} {\bf B499} (1997) 673--689,
  [\href{http://xxx.lanl.gov/abs/hep-th/9701109}{{\tt hep-th/9701109}}].

\bibitem{Miramontes:1999hx}
J.~Miramontes and C.~Fernandez-Pousa, {\it {Integrable quantum field theories
  with unstable particles}},  {\em Phys. Lett.} {\bf B472} (2000) 392--401,
  [\href{http://xxx.lanl.gov/abs/hep-th/9910218}{{\tt hep-th/9910218}}].

\bibitem{Dorey:2004qc}
P.~Dorey and J.~Miramontes, {\it {Mass scales and crossover phenomena in the
  homogeneous sine-Gordon models}},  {\em Nucl. Phys.} {\bf B697} (2004)
  405--461, [\href{http://xxx.lanl.gov/abs/hep-th/0405275}{{\tt
  hep-th/0405275}}].

\bibitem{CastroAlvaredo:2001ih}
O.~A. Castro-Alvaredo, {\it {Bootstrap methods in 1+1 dimensional quantum field
  theories: The Homogeneous Sine-Gordon models}},
  \href{http://xxx.lanl.gov/abs/hep-th/0109212}{{\tt hep-th/0109212}}.

\bibitem{Lund:1976ze}
F.~Lund and T.~Regge, {\it {Unified approach to strings and vortices with
  soliton solutions}},  {\em Phys. Rev.} {\bf D14} (1976) 1524.

\bibitem{Lund:1977dt}
F.~Lund, {\it {Example of a relativistic, completely integrable, hamiltonian
  system}},  {\em Phys. Rev. Lett.} {\bf 38} (1977) 1175.

\bibitem{Getmanov:1977hk}
B.~Getmanov, {\it {New Lorentz invariant systems with exact multi-soliton
  solutions}},  {\em JETP Lett.} {\bf 25} (1977) 119.

\bibitem{Destri:1989}
C.~Destri and H.~de~Vega, {\it {Light-cone lattices and the exact solution of
  chiral fermion and sigma models}},  {\em J. Phys. A: Math. Gen.} {\bf 22}
  (1989) 1329.

\bibitem{Caetano:2010zd}
J.~Caetano, {\it {Unified approach to the SU(2) principal chiral field model at
  finite volume}},  \href{http://xxx.lanl.gov/abs/1012.2600}{{\tt
  arXiv:1012.2600}}.

\bibitem{Falceto:1992bf}
F.~Falceto and K.~Gawedzki, {\it {Lattice Wess-Zumino-Witten model and quantum
  groups}},  {\em J. Geom. Phys.} {\bf 11} (1993) 251--279,
  [\href{http://xxx.lanl.gov/abs/hep-th/9209076}{{\tt hep-th/9209076}}].

\bibitem{Alekseev:1991wq}
A.~Alekseev, L.~Faddeev, M.~Semenov-Tian-Shansky, and A.~Volkov, {\it {The
  Unraveling of the quantum group structure in the WZNW theory}},
  Preprint CERN-TH-5981/91 (1991).

\bibitem{Alekseev:1992wn}
A.~Alekseev, L.~Faddeev, and M.~Semenov-Tian-Shansky, {\it {Hidden quantum
  groups inside Kac-Moody algebra}},  {\em Commun. Math. Phys.} {\bf 149}
  (1992) 335--345.

\bibitem{Dorey:1994mg}
N.~Dorey and T.~J. Hollowood, {\it {Quantum scattering of charged solitons in
  the complex sine-Gordon model}},  {\em Nucl. Phys.} {\bf B440} (1995)
  215--236, [\href{http://xxx.lanl.gov/abs/hep-th/9410140}{{\tt
  hep-th/9410140}}].

\bibitem{Hoare:2010fb}
B.~Hoare and A.~Tseytlin, {\it {On the perturbative S-matrix of generalized
  sine-Gordon models}},  {\em JHEP} {\bf 1011} (2010) 111,
  [\href{http://xxx.lanl.gov/abs/1008.4914}{{\tt arXiv:1008.4914}}].

\bibitem{deVega:1981ka}
H.~de~Vega and J.~Maillet, {\it {Renormalization character and quantum S matrix
  for a classically integrable theory}},  {\em Phys. Lett.} {\bf B101} (1981)
  302.

\bibitem{deVega:1982sh}
H.~de~Vega and J.~Maillet, {\it {Semiclassical quantization of the complex
  sine-Gordon field theory}},  {\em Phys. Rev.} {\bf D28} (1983) 1441.

\bibitem{Maillet:1981pj}
J.~Maillet, {\it {Quantum $U(1)$ invariant theory from integrable classical
  models}},  {\em Phys. Rev.} {\bf D26} (1982) 2755.

\bibitem{Maldacena:1997re}
J.~M. Maldacena, {\it {The Large N limit of superconformal field theories and
  supergravity}},  {\em Adv. Theor. Math. Phys.} {\bf 2} (1998) 231--252,
  [\href{http://xxx.lanl.gov/abs/hep-th/9711200}{{\tt hep-th/9711200}}].

\bibitem{Gubser:1998bc}
S.~Gubser, I.~R. Klebanov, and A.~M. Polyakov, {\it {Gauge theory correlators
  from noncritical string theory}},  {\em Phys. Lett.} {\bf B428} (1998)
  105--114, [\href{http://xxx.lanl.gov/abs/hep-th/9802109}{{\tt
  hep-th/9802109}}].

\bibitem{Witten:1998qj}
E.~Witten, {\it {Anti-de Sitter space and holography}},  {\em Adv. Theor. Math.
  Phys.} {\bf 2} (1998) 253--291,
  [\href{http://xxx.lanl.gov/abs/hep-th/9802150}{{\tt hep-th/9802150}}].

\bibitem{Beisert:2010jr}
N.~Beisert and al., {\it {Review of AdS/CFT Integrability: An Overview}},  {\em
  Lett. Math. Phys.} {\bf 99} (2012) 3,
  [\href{http://xxx.lanl.gov/abs/1012.3982}{{\tt arXiv:1012.3982}}].

\bibitem{Magro:2008dv}
M.~Magro, {\it {The classical exchange algebra of $AdS_5 \times S^5$ string
  theory}},  {\em JHEP} {\bf 0901} (2009) 021,
  [\href{http://xxx.lanl.gov/abs/0810.4136}{{\tt arXiv:0810.4136}}].

\end{thebibliography}
\end{document}